\documentclass[epj]{webofc}
\usepackage[varg]{txfonts}   
\usepackage{color}
\usepackage[utf8]{inputenc}
\definecolor{dgreen}{cmyk}{1.,0.,1.,0.2}        
\definecolor{orange}{cmyk}{0.,0.353,1.,0.}    

\def\pt{p_{\rm T}}

\wocname{EPJ Web of Conferences}
\woctitle{ICNFP 2016}
\begin{document}
\selectlanguage{english}
\title{Overview of recent azimuthal correlation measurements from ALICE}
%
%

\author{You Zhou (for the ALICE Collaboration) \inst{1}\fnsep\thanks{\email{you.zhou@cern.ch}} }

\institute{
Niels Bohr Institute, University of Copenhagen, Blegdamsvej 17, 2100 Copenhagen, Denmark
}

\abstract{%
 Azimuthal correlations are a powerful tool to probe the properties and the evolution of the collision system. In this proceedings, we will review the recent azimuthal correlation measurements from ALICE at the LHC. The comparison to other experimental measurements and various theoretical calculations will be discussed as well. }
\maketitle
\section{Introduction}
\label{intro}

One of the fundamental questions in the phenomenology of Quantum Chromo Dynamics (QCD) 
is what are the properties of matter at extreme densities and temperatures where quarks
and gluons are in a new state of matter, the so-called Quark Gluon Plasma (QGP)~\cite{Lee:1978mf, Shuryak:1980tp}. 
For a non-central collision between two nuclei of lead, the initial nuclear overlap region is spatially
asymmetric with an ``almond-like'' shape.
The translation of such anisotropies in the fluctuating initial conditions into anisotropies in momentum space would not occur if individual nucleon-nucleon collisions emitted independently of each other. 
Thus, the collective expansion of matter created in collisions of heavy-ions, refereed to phenomenon of ``collective flow'', proved to be one of the best probes to study the detailed properties of these unknown states of matter. Measurements of the collective expansion of the QGP, in particular the azimuthal anisotropies in the expansion, provide us with constraints on the fluctuating initial conditions and the properties of the system such as the equation of state (EoS) and transport coefficients like shear viscosity over entropy density ratio $\eta/s$. 
The characterization of an anisotropic ``shape'' in the final momentum space is typically performed using a Fourier decomposition of the azimuthal angle distribution of final state particles $P(\varphi)$~\cite{Ollitrault:1992bk, Voloshin:1994mz}:
\begin{equation}
P(\varphi) = \frac{1}{2\pi} \sum_{n=-\infty}^{+\infty} {\overrightarrow{V_{n}} \, e^{-in\varphi} }
\end{equation}
where $\varphi$ is the azimuthal angle of emitted particles, $\overrightarrow{V_{n}}$ is the $n$-th order flow-vector defined as $\overrightarrow{V_{n}} = v_{n}\,e^{in\Psi_{n}}$, its magnitude $v_{n}$ is the $n$-th order anisotropic flow harmonic and its orientation is symmetry plane (participant plane) angle $\Psi_{n}$. Alternatively, this anisotropy can be generally given by the joint probability density function ($p.d.f.$) in terms of $v_{n}$ and $\Psi_{n}$ as:
\begin{equation}
P(v_{m}, v_{n}, ..., \Psi_{m}, \Psi_{n}, ...) = \frac{1}{N_{event}} \frac{dN_{event}} {v_{m} v_{n} \cdot \cdot \cdot {\rm d} v_{m} \,{\rm d} v_{n} \cdot \cdot \cdot {\rm d}\Psi_{m} \, {\rm d}\Psi_{n} \cdot \cdot \cdot}.
\end{equation}

Anisotropic flow was a major piece of the RHIC program~\cite{Ackermann:2000tr, Adcox:2002ms, Back:2002gz}, and its characterization in terms of hydrodynamics~\cite{Huovinen:2001cy, Kolb:2000fha, Luzum:2008cw, Song:2007ux, Song:2010mg} was taken as one of the most important evidences of observation of the strongly-coupled quark-gluon plasma (sQGP) at RHIC.
The precision measurements of anisotropic flow based on the huge data collected at the LHC experiments~\cite{Aamodt:2010pa, ALICE:2011ab, ATLAS:2011ah, Chatrchyan:2012wg} and the successful descriptions by hydrodynamic calculations, demonstrate that the QGP created at the LHC continually behaves like a strongly coupled liquid with a very small $\eta/s$~\cite{Heinz:2013th, Luzum:2013yya, Huovinen:2013wma, Shuryak:2014zxa, Song:2013gia, Dusling:2015gta} which is close to a quantum limit 1/4$\pi$~\cite{Kovtun:2004de}. 

The anisotropic flow measurements are recently extended by ALICE Collaboration to the forward pseudorapidity region with its unique coverage ($-3.5 < \eta < 5$)~\cite{Adam:2016ows}, to identified particles with its powerful particle identifications~\cite{Adam:2016nfo}. In addition, the correlations between different order flow harmonics with newly proposed symmetric cumulants are investigated in Pb--Pb collisions at $\sqrt{s_{_{\rm NN}}} = 2.76$ TeV~\cite{ALICE:2016kpq}. Furthermore, based on the data taken in recent LHC Run2 program, the anisotropic flow of charged particle are measured at the highest collision energy so far at $\sqrt{s_{_{\rm NN}}} = 5.02$ TeV Pb--Pb collisions~\cite{Adam:2016izf}. All the measurements provide crucial information on the initial conditions and the dynamic of the system which were poorly known before.
In this proceedings we will review recent ALICE measurements of anisotropic flow at the LHC.

\section{Forward flow}
\label{sec-1}

\begin{figure}[h]
\begin{center}
\includegraphics[width=12cm]{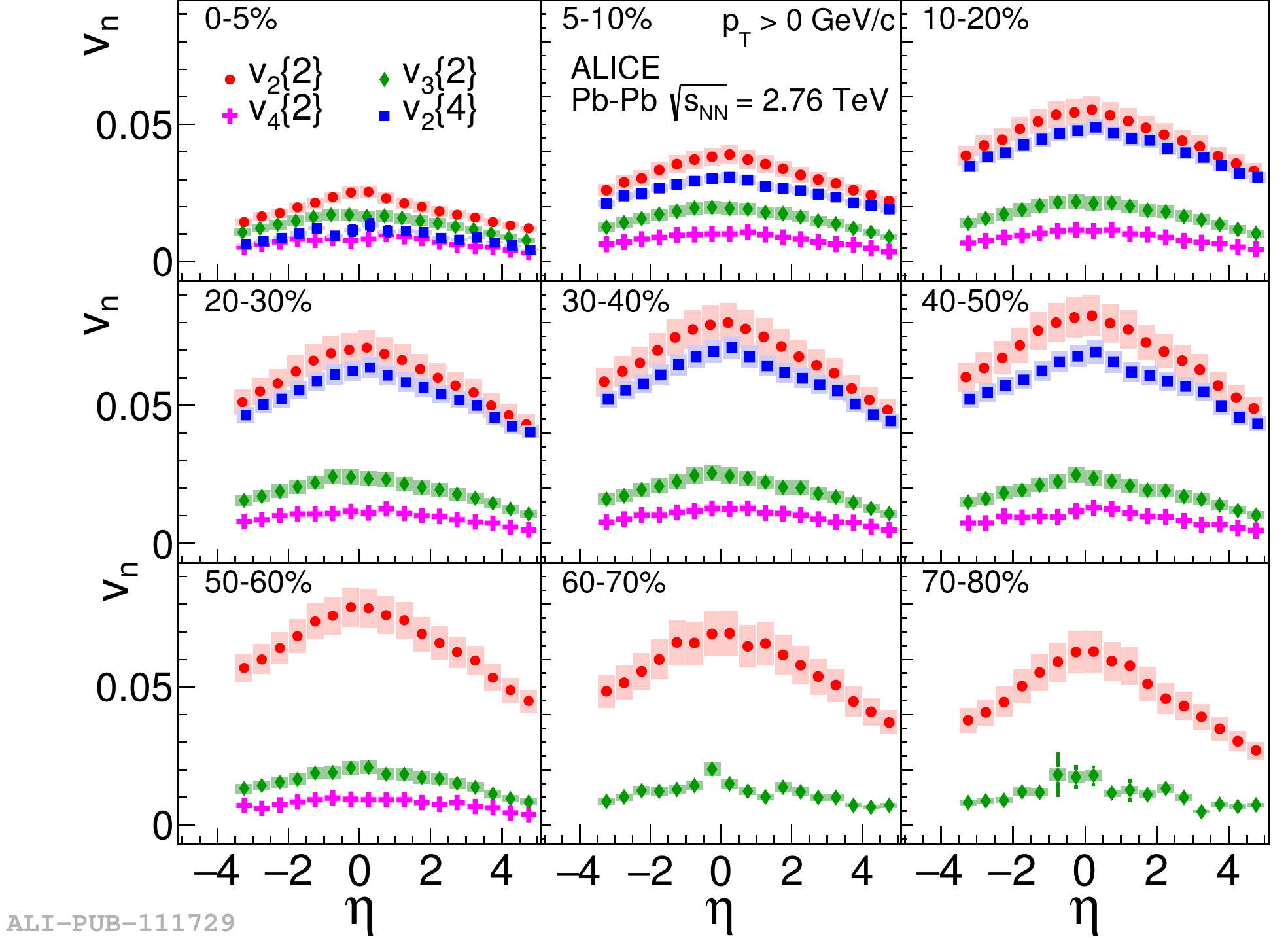}
\caption{Pseudorapidity dependence of $v_2$, $v_3$ and $v_4$ in various centrality classes. Figure is taken from~\cite{Adam:2016ows}.}
\label{fig-10}      
\end{center}
\end{figure}

The pseudorapidity dependence of $v_{2}$ over a wide range ($-5.0 < \eta < 5.3$) and various of collision energies, and systems was investigated by PHOBOS Collaboration~\cite{Back:2002gz, Alver:2010gr, Back:2004zg, Alver:2006wh}. It was reported that in the rest frame of one of the colliding nuclei ($\eta-y_{\rm beam}$), $v_{2}$ is independent of collision energy. Similar feature was observed in multiplicity density distributions~\cite{Bearden:2001qq}. They suggest that particle production is independent of the collision energy at forward rapidity, the fragmentation region. Such effect is known as extended longitudinal scaling. 
Not only $v_{2}$ but also $v_{3}$ and $v_{4}$ have been measured over the widest $\eta$-range at the LHC, in Pb--Pb collisions at $\sqrt{s_{_{\rm NN}}} = 2.76$ TeV using the ALICE detector~\cite{Adam:2016ows}. 
The pseudorapidity dependence of anisotropic flow in nine centrality classes is shown in Fig.~\ref{fig-10}. First of all, it is seen that $v_{2}$ measured with both 2- and 4-particle cumulants increases from 0-5\% to 40-50\% centrality classes, $v_{2}\{2\}$ decreases slightly in more peripheral collisions over the entire pseudorapidity range. Compared to $v_{2}$, the higher harmonic flow $v_{3}$ and $v_{4}$ show weaker centrality dependence. This might be due to the fact that initial state fluctuations play a prominent role for higher harmonics, as the centrality dependence of the corresponding eccentricities $\varepsilon_{n}$ (with $n \geq 3$) are more modest relative to $\varepsilon_2$.

\begin{figure}[h]
\centering
\includegraphics[width=6.7cm,clip]{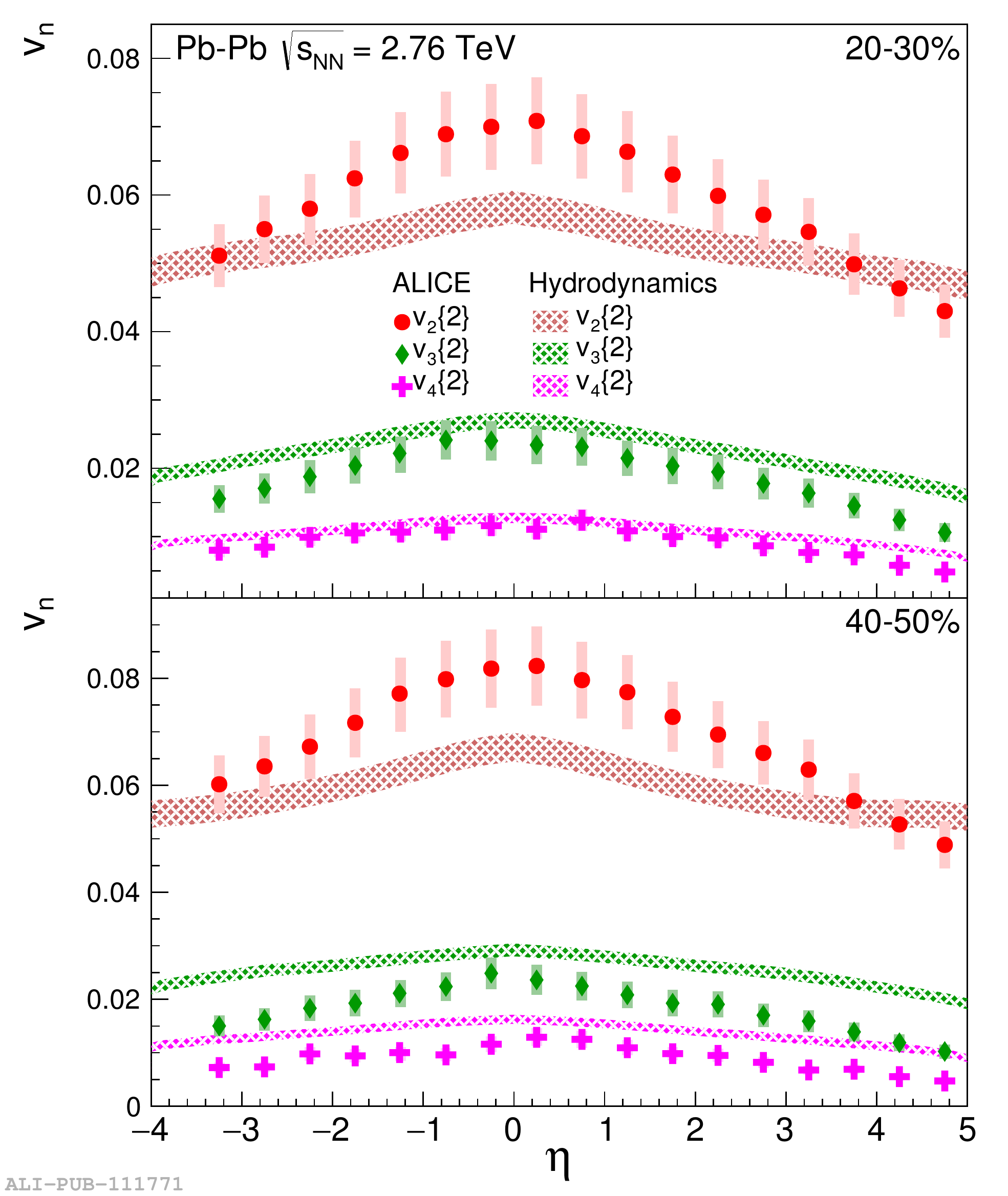}
\includegraphics[width=6.3cm,clip]{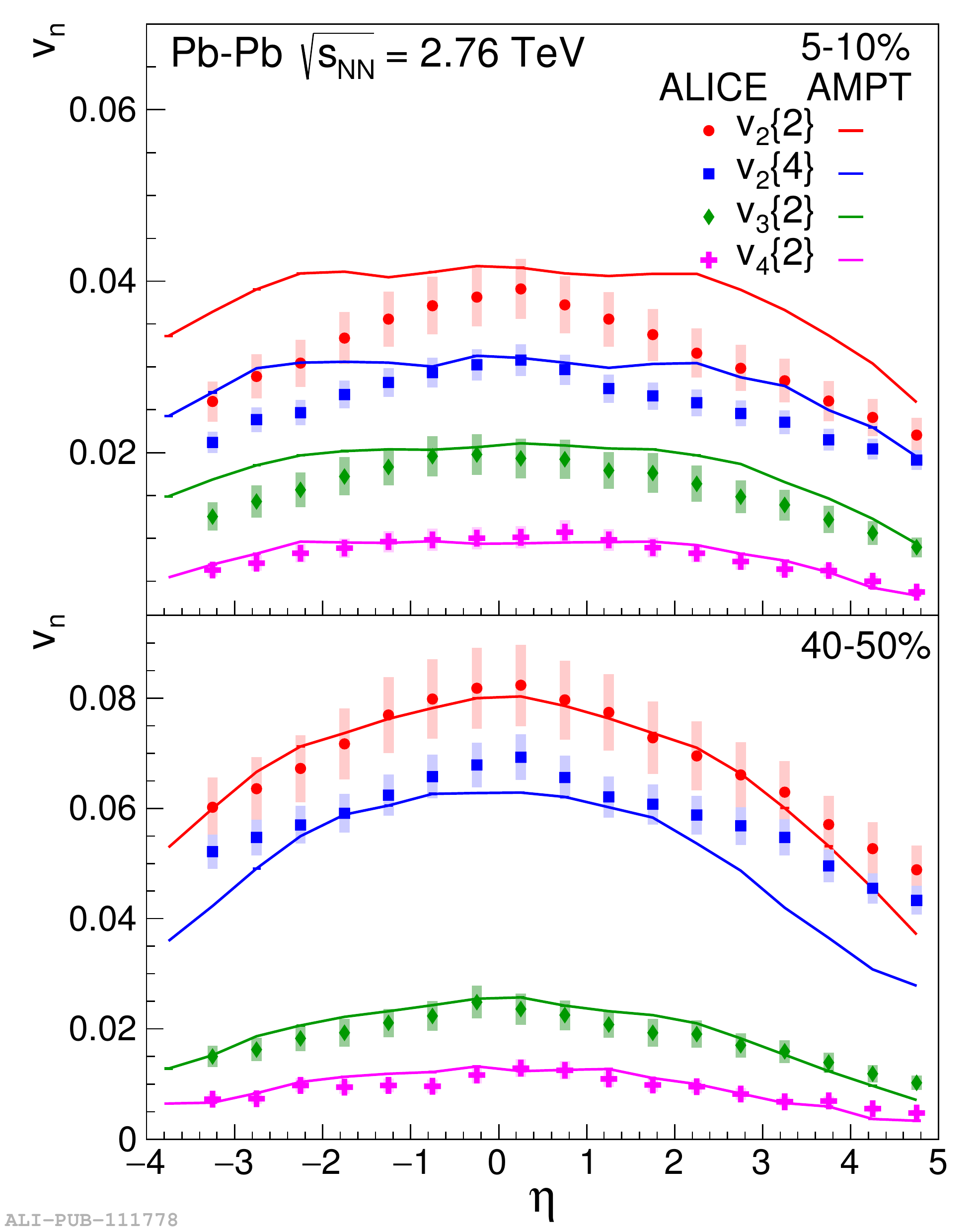}
\caption{Pseudorapidity dependence of $v_2$, $v_3$ and $v_4$ in selected centrality class. Calculations from hydrodynamic (left) and AMPT (right) are presented for comparisons~\cite{Denicol:2015nhu,Xu:2011fi}. Figure is taken from~\cite{Adam:2016ows}.}
\label{fig-11}      
\end{figure}

\begin{figure}[h]
\centering
\includegraphics[width=6.5cm,clip]{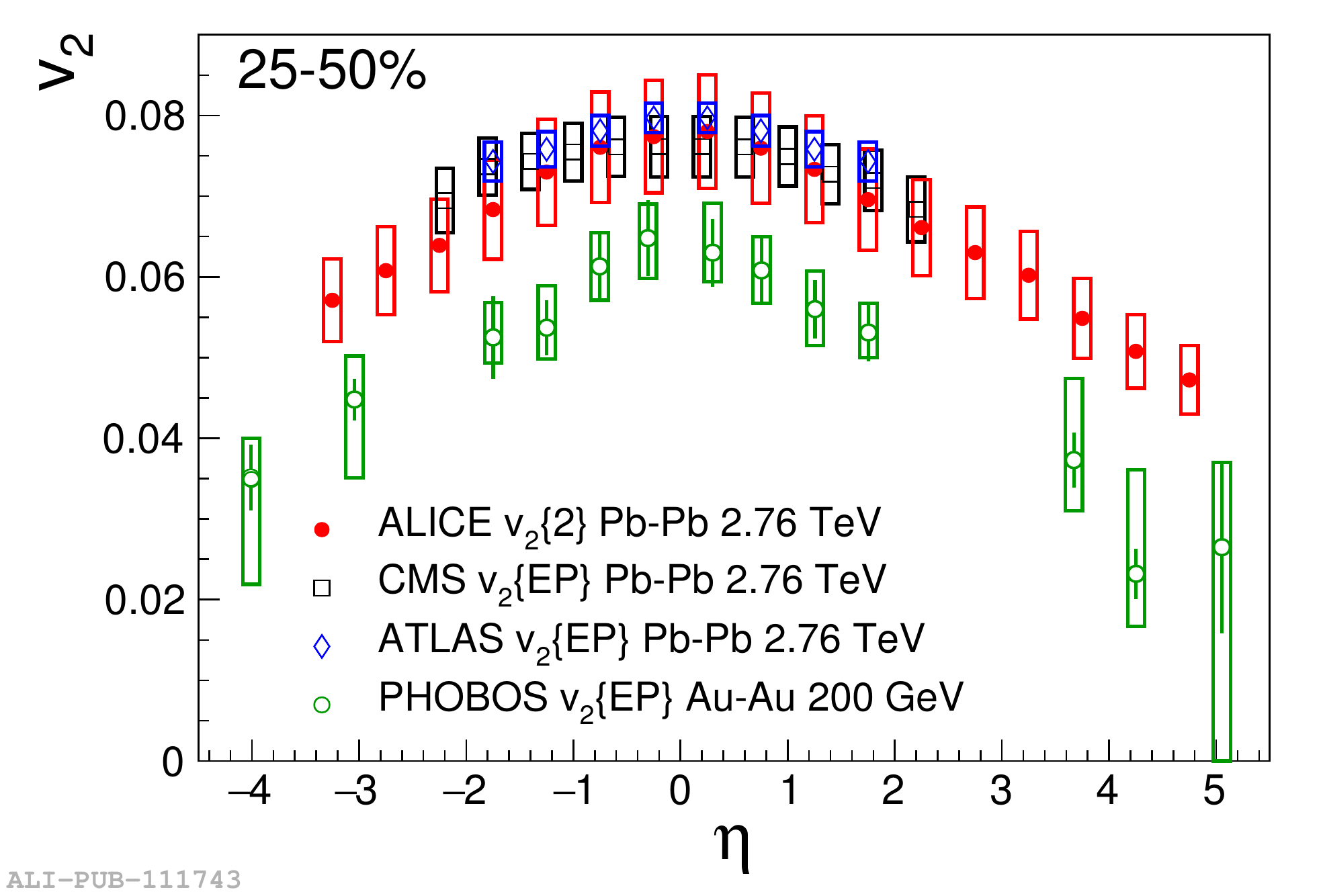}
\includegraphics[width=6.5cm,clip]{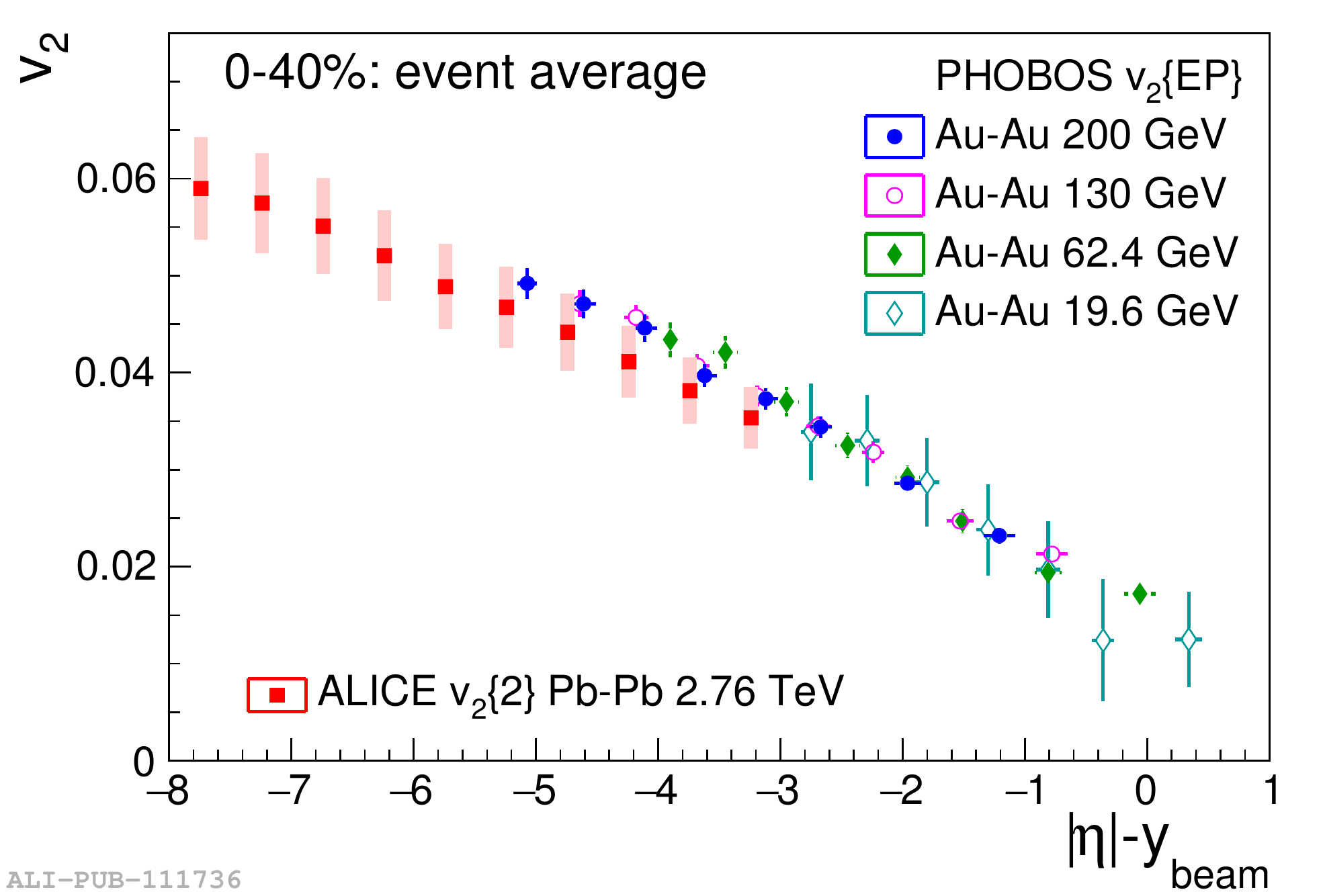}
\caption{$v_2(\eta)$ measured at RHIC~\cite{Back:2002gz, Alver:2010gr, Back:2004zg} and the LHC~\cite{Aad:2014eoa,Chatrchyan:2012ta} (left); observed $v_{2}$ in the rest frame of the projectiles for the event averaged 0-40 \% centrality range (right). Figure is taken from~\cite{Adam:2016ows}.}
\label{fig-12}      
\end{figure}

\begin{figure}[h]
\centering
\includegraphics[width=6.3cm,clip]{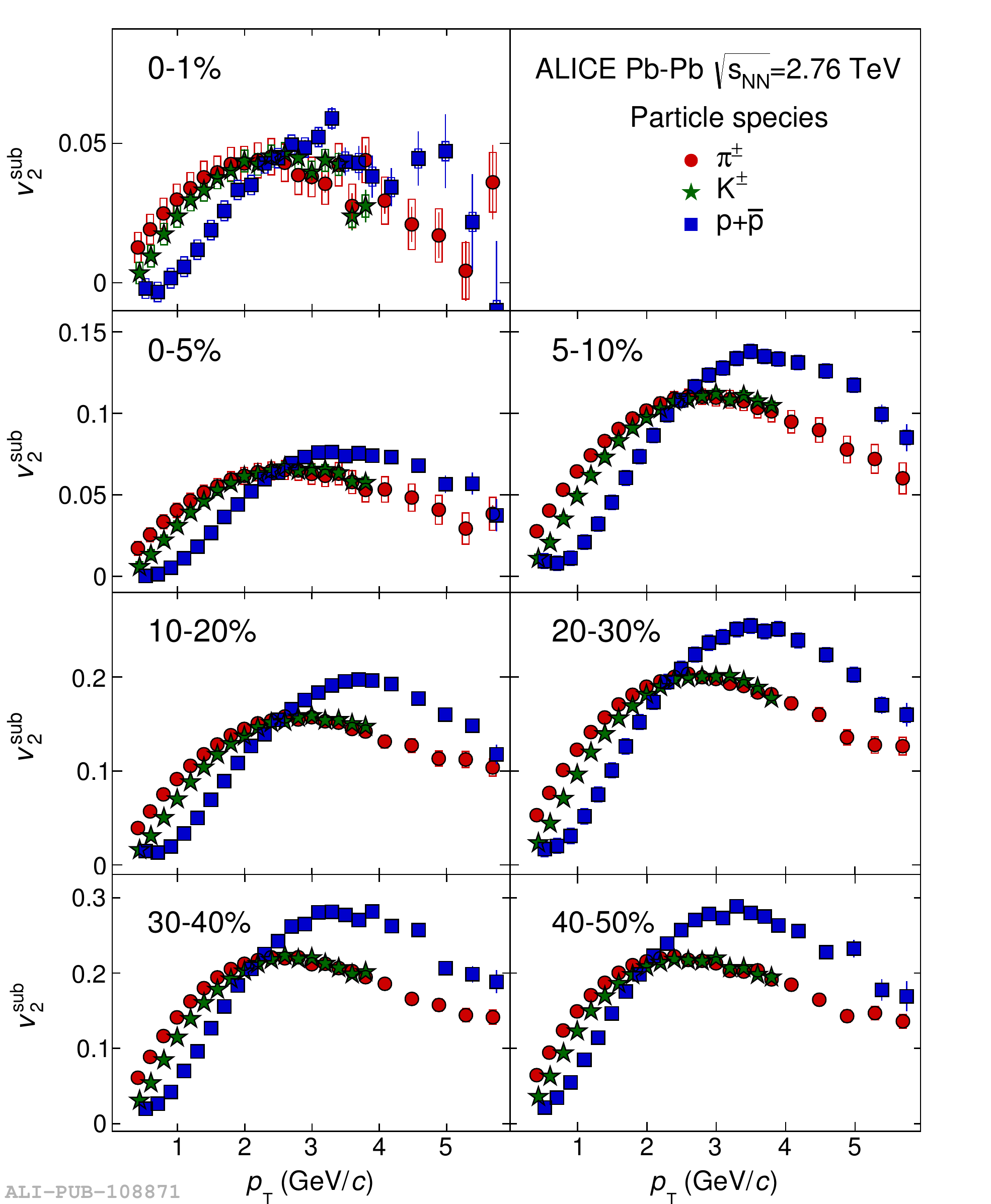}
\includegraphics[width=6.3cm,clip]{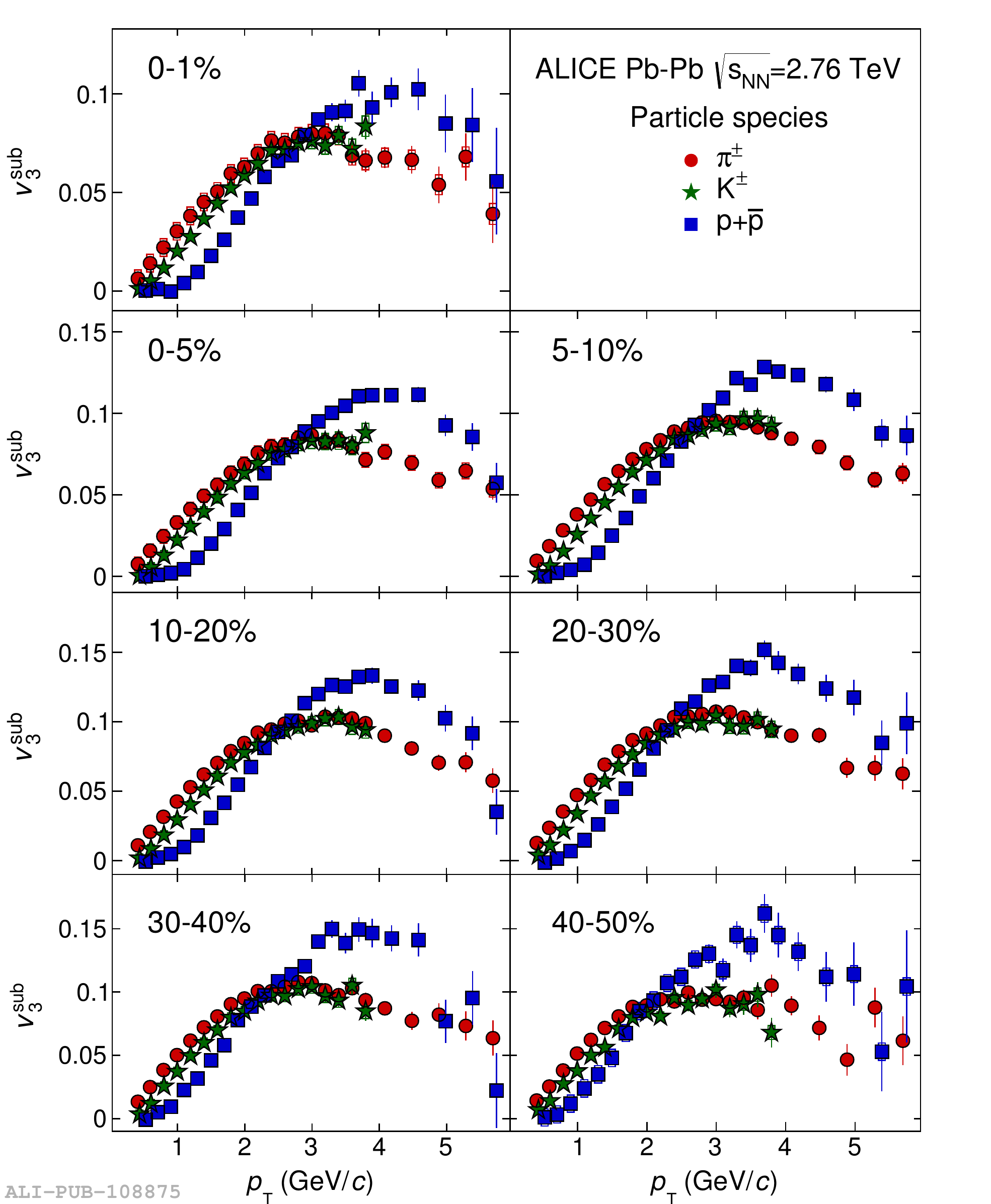}
\includegraphics[width=6.3cm,clip]{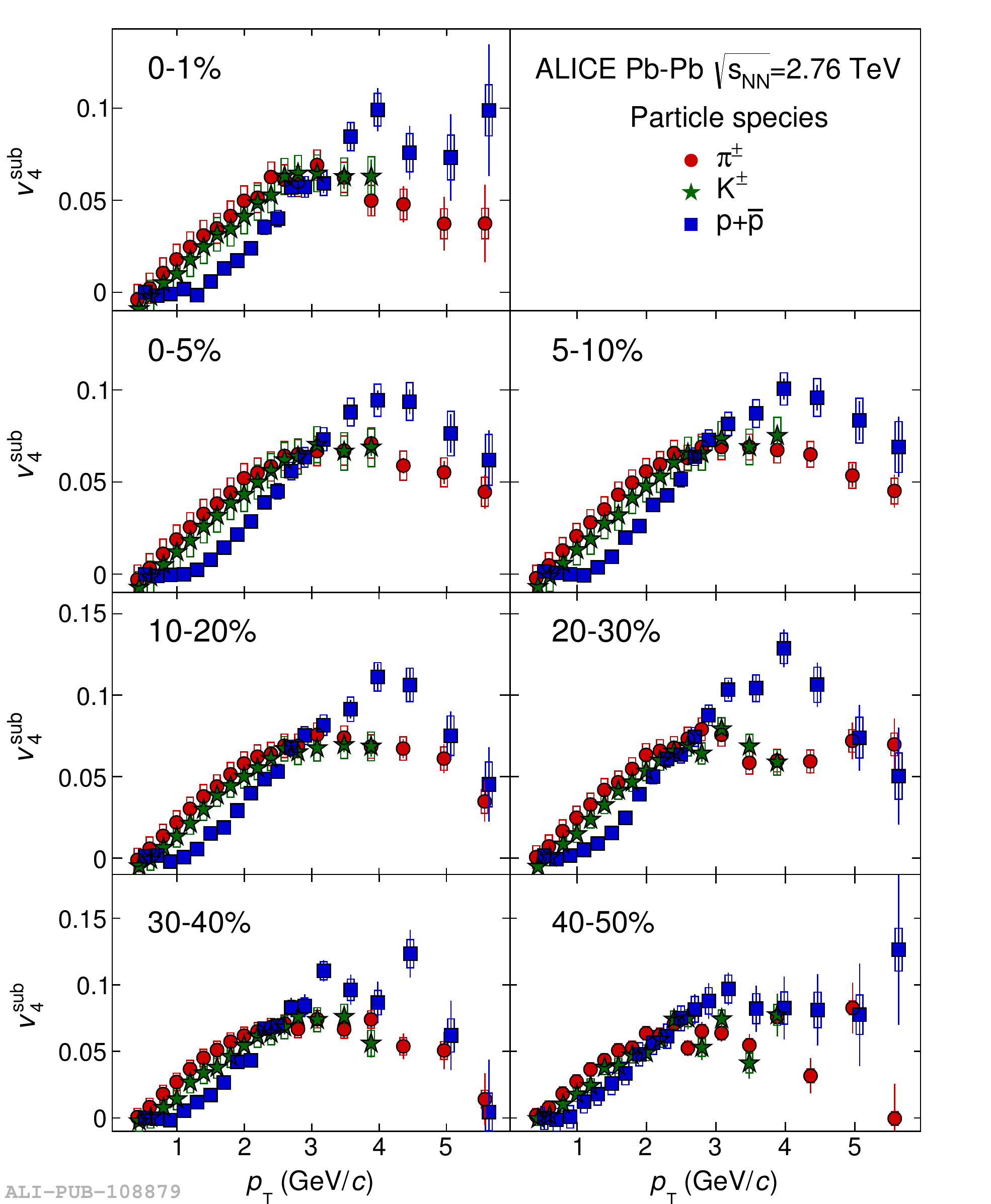}
\includegraphics[width=6.3cm,clip]{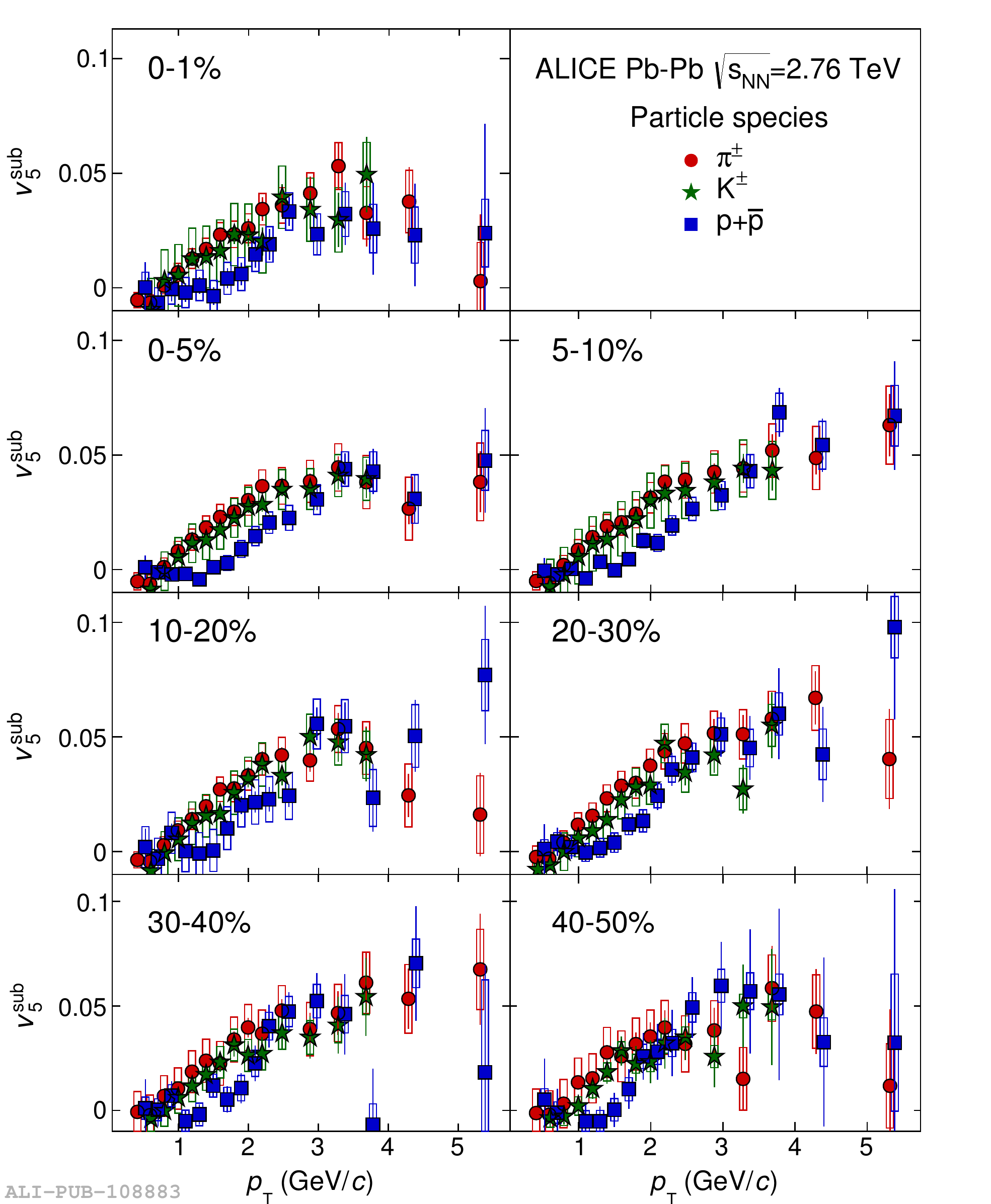}
\caption{The $p_{\rm T}$-differential $v_{2}^{\rm sub}$ (top left), $v_{3}^{\rm sub}$ (top right), $v_{4}^{\rm sub}$ (bottom left) and $v_{5}^{\rm sub}$ (bottom right) for identified particle species in various centrality classes in Pb--Pb collisions at $\sqrt{s_{\rm NN}} =$ 2.76 TeV. Figure is taken from~\cite{Adam:2016nfo}.}
\label{fig-20}      
\end{figure}

In order to understand the role of $\eta/s$ and hadronic interactions in forward pseudorapidity region, the measurements are compared to viscous hydrodynamic calculations~\cite{Denicol:2015nhu} whose input parameters e.g. temperature dependent shear viscosity over entry density ratio $\eta/s({\rm T})$ are tuned to fit the RHIC data, shown in Fig.~\ref{fig-11} (left). Although it can nicely describe the PHOBOS measurements of $v_{2}(\eta)$~\cite{Back:2002gz, Alver:2010gr, Back:2004zg}, it clearly fails to quantitatively describe the LHC data using the same parameterization. More specifically, for selected two centralities, hydrodynamic calculations underestimates the measured $v_{2}$ while they overestimate the measurements of $v_{3}$ and $v_4$. Further tuning of parameterization of $\eta/s({\rm T})$ and/or the modification of initial conditions are necessary to describe the data better. At the same time, A multiphase transport (AMPT) model~\cite{Xu:2011fi} with a string melting scenario is used for comparisons. This is a hybrid transport model consists of four main stages: initial conditions, partonic interactions, hadronization, and hadronic rescattering. It could reproduce compatible $p_{\rm T}$ integrated $v_{2}$, $v_{3}$ and $v_{4}$ as data at the LHC~\cite{Xu:2011fi}. Good agreements of anisotropic flow from AMPT and data are observed in Fig.~\ref{fig-11} (right) in 40-50\% centrality class, except $v_{2}\{4\}$ at forward pseudorapidity. In more central collisions, AMPT overestimates $v_{2}$ and $v_{3}$ nevertheless agrees with $v_{4}$ measurements.

\begin{figure}[h]
\centering
\includegraphics[width=6.5cm,clip]{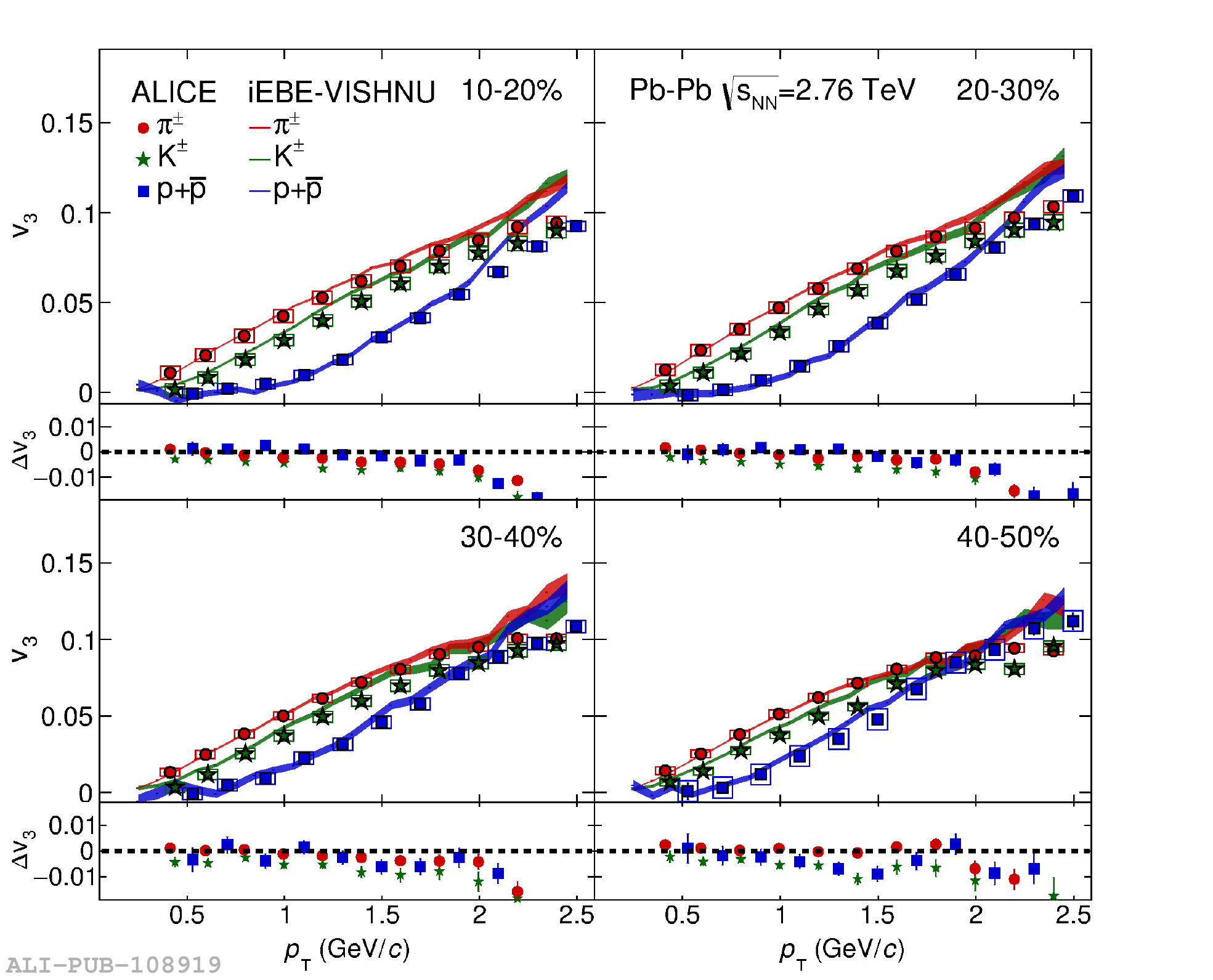}
\includegraphics[width=6.5cm,clip]{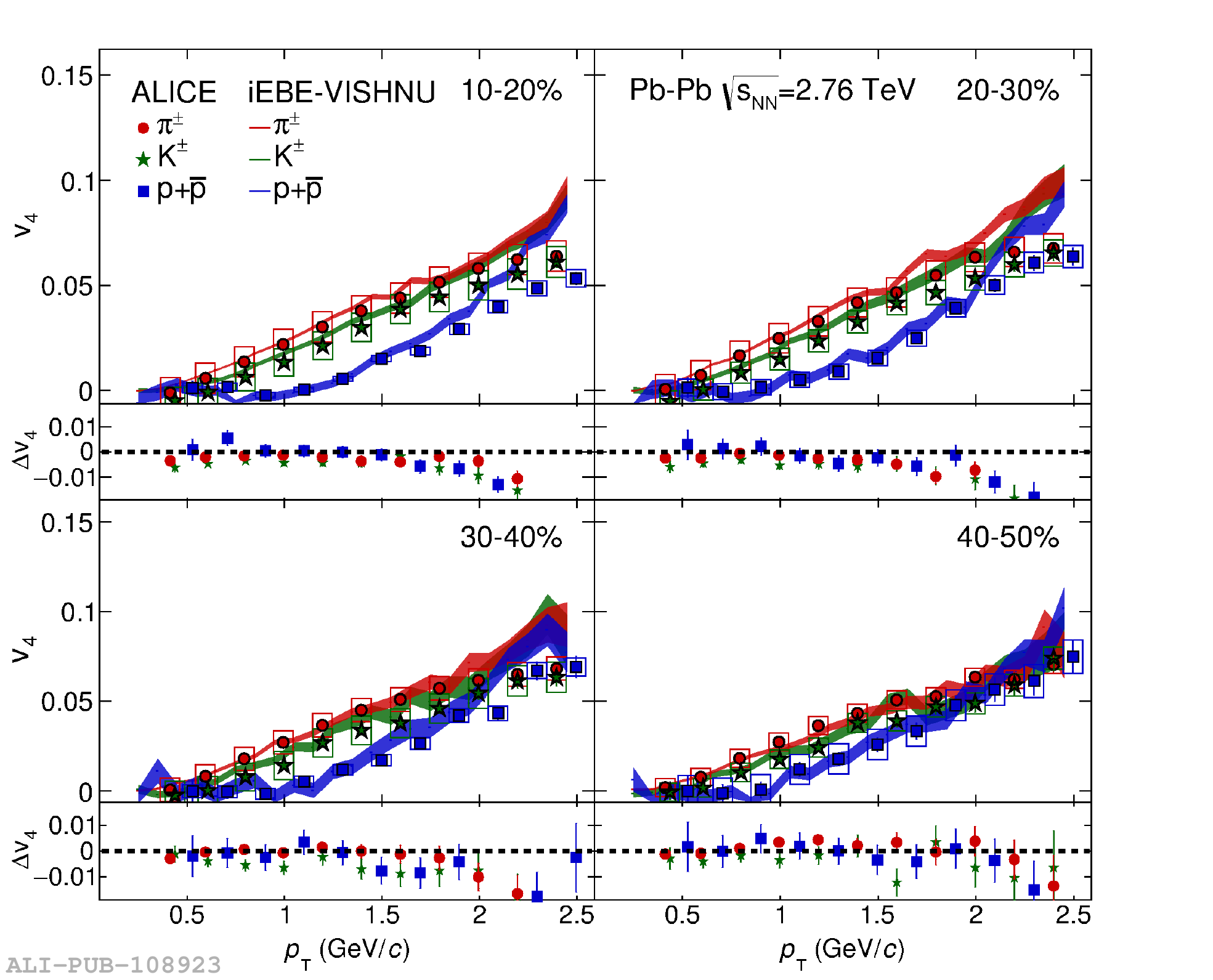}
\caption{$p_{\rm T}$-differential $v_{3}^{\rm sub}$ (left) and $v_{4}^{\rm sub}$ (right) for identified particle species in various centrality classes in Pb--Pb collisions at $\sqrt{s_{\rm NN}} =$ 2.76 TeV. Comparisons to iEBE-VISHNU calculations~\cite{Xu:2016hmp} are presented. Figure is taken from~\cite{Adam:2016nfo}.}
\label{fig-21}       
\end{figure}

The pseudorapidity dependence of $v_{2}$ was measured both at RHIC and other LHC experiments before, comparisons to ALICE measurements are shown in Figure~\ref{fig-12} (left) with a wider centrality range 25-50 \%. It is found that for the overlapped pseudorapidity region, the recent $v_{2}(\eta)$ measurements by ALICE is consistent with published results from ATLAS~\cite{Aad:2014eoa} and CMS~\cite{Chatrchyan:2012ta} Collaborations within the systematic uncertainties. All the results at the LHC are systematically above the previous $v_{2}(\eta)$ from PHOBOS at RHIC. As mentioned above, it was reported that in the rest frame of one of the colliding nuclei ($\eta - y_{beam}$), $v_{2}$ is energy independent. This is examined also at the LHC as shown in Fig.~\ref{fig-12} (right). It is seen that the extended longitudinal scaling is found to hold in Au--Au collisions at 19.6 GeV up to 2.76 TeV in Pb--Pb collisions, similar conclusion was obtained by ATLAS and CMS with smaller pseudorapidity coverage~\cite{Aad:2014eoa,Chatrchyan:2012ta}.

\section{Anisotropic flow of identified particles}
\label{sec-2}

The measurements of anisotropic flow for charged particles provided a strong testing ground for hydrodynamical
calculations that attempts to describe the dynamical evolution of the system created in heavy-ion collisions.
In addition, the measurements for identified particles, together with the comparisons to various theoretical calculations, can help us further constrain initial conditions and dynamics of the collisions, i.e. the properties of the system (e.g. $\eta/s$) as well as probe the particle production mechanism (e.g. coalescence).
Using the unique capability of particle identification of ALICE detector, not only the $v_{2}$ but also higher harmonic flow $v_{3}$, $v_{4}$ and $v_{5}$ are measured in Pb--Pb collisions at $\sqrt{s_{\rm NN}} =$ 2.76 TeV~\cite{Abelev:2014pua,Adam:2016nfo}. 
It is observed in Fig.~\ref{fig-20} (top left) that the $p_{\rm T}$-differential $v_{2}$ for charged pions, kaons and protons show an obvious mass dependence feature for $\pt < 2-3$ GeV/$c$, in all presented centrality classes. This mass ordering originates from the fact that radial flow creates a depletion in the particle spectrum at low $\pt$, which increases with increasing particle mass and transverse velocity. This depletion leads to heavier particles (e.g. protons) having smaller $v_{2}$ values compared to lighter ones (e.g. pions) at given values of $\pt$. 
A crossing of identified particles $v_2$ is observed around $\pt \approx$ 2.5 GeV/$c$ in the most central collisions and happens at lower $\pt$ values in more peripheral collisions, agree with what reported before~\cite{Abelev:2014pua}.
Similar to $v_{2}$, a clear mass ordering feature is seen in higher harmonic flow $v_{3}$, $v_4$ and $v_{5}$ of charged pions, kaons and protons for $\pt<2-3$ GeV/$c$. This could be understood as the interplay between the radial flow and anisotropic flow. Furthermore, the crossing of baryon and meson $v_{n}$ is also observed. The $\pt$ region where this crossing happens depending on on the centrality and the order of flow harmonic $n$.

\begin{figure}[h]
\centering
\includegraphics[width=6.3cm,clip]{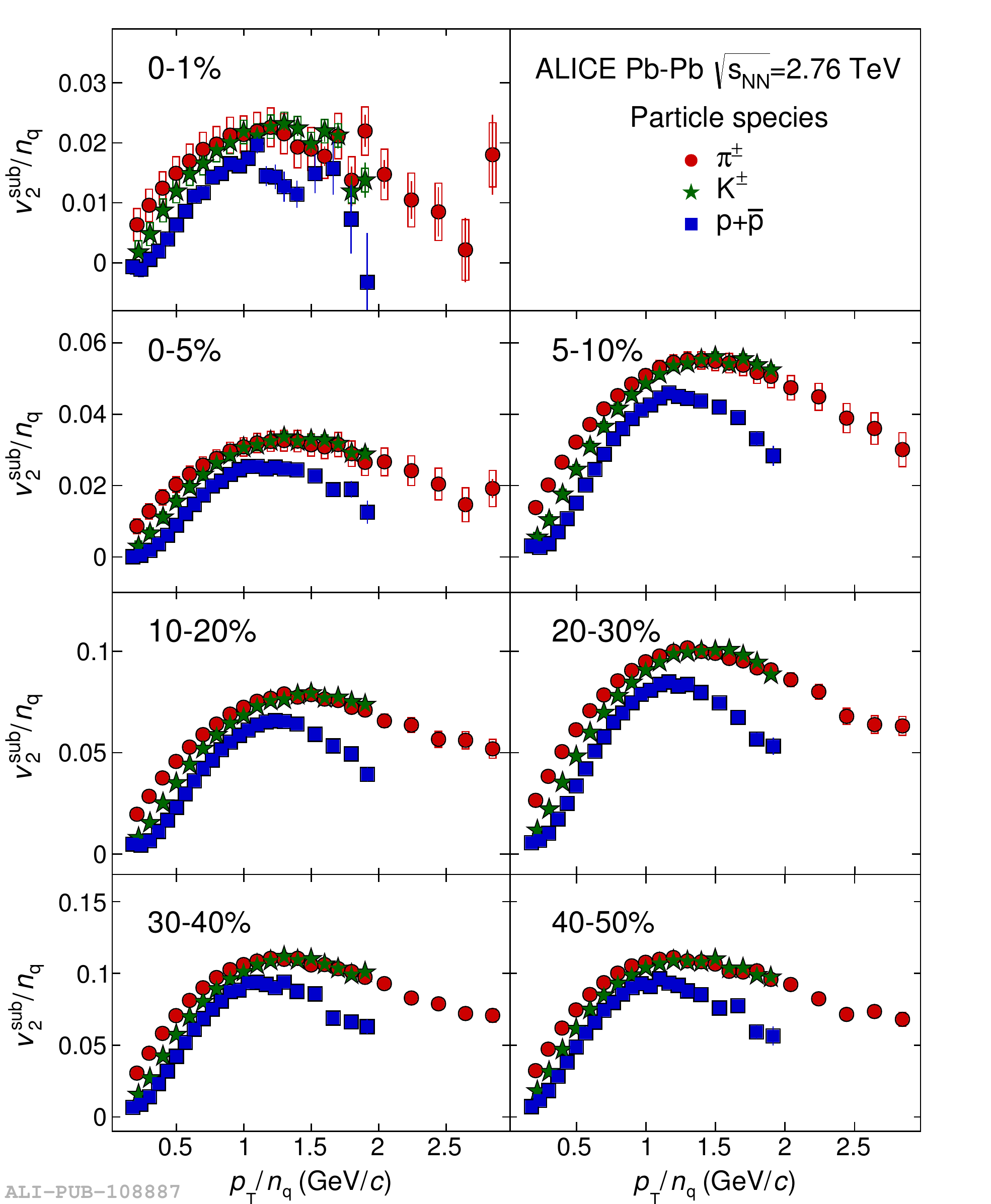}
\includegraphics[width=6.3cm,clip]{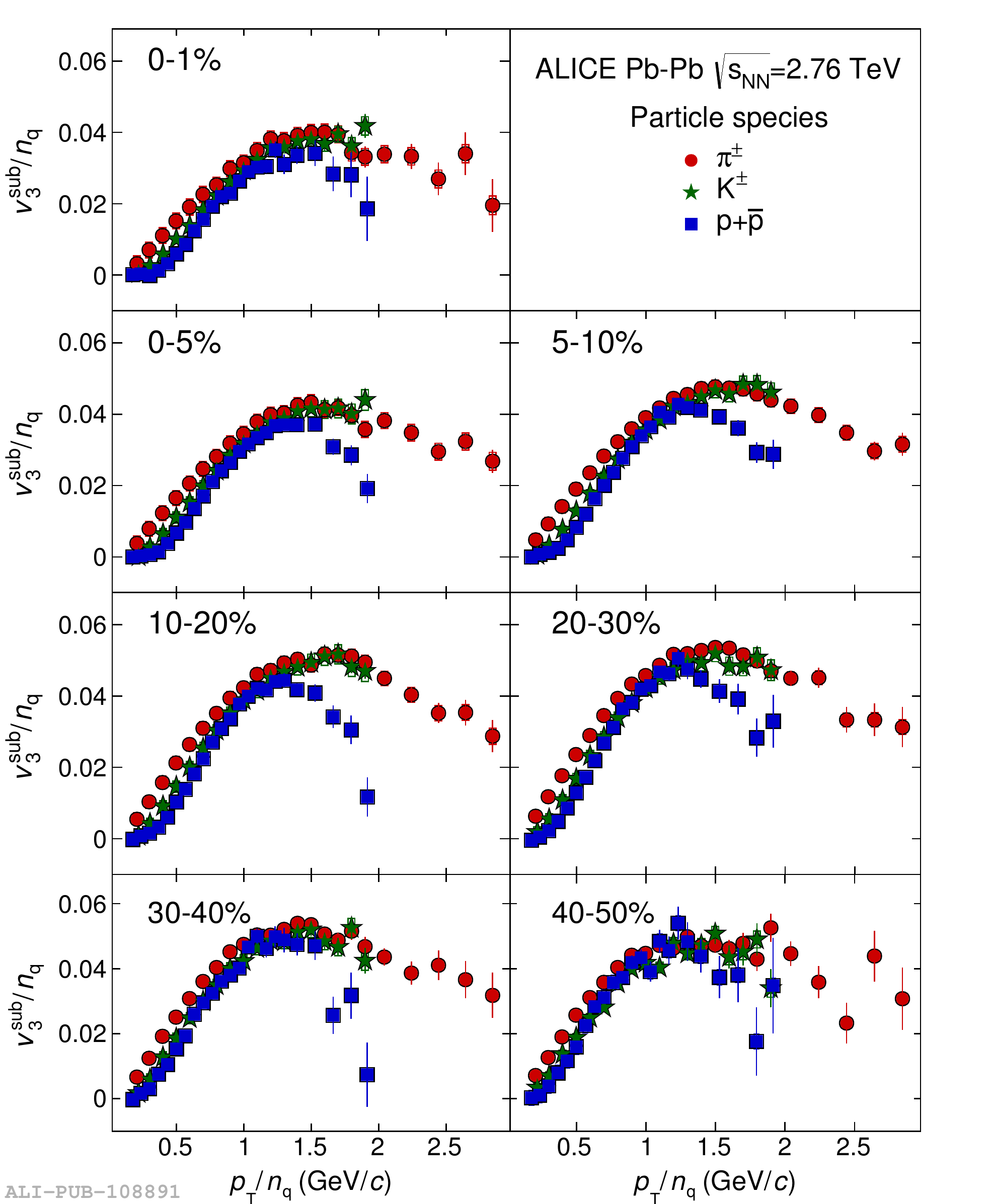}
\includegraphics[width=6.3cm,clip]{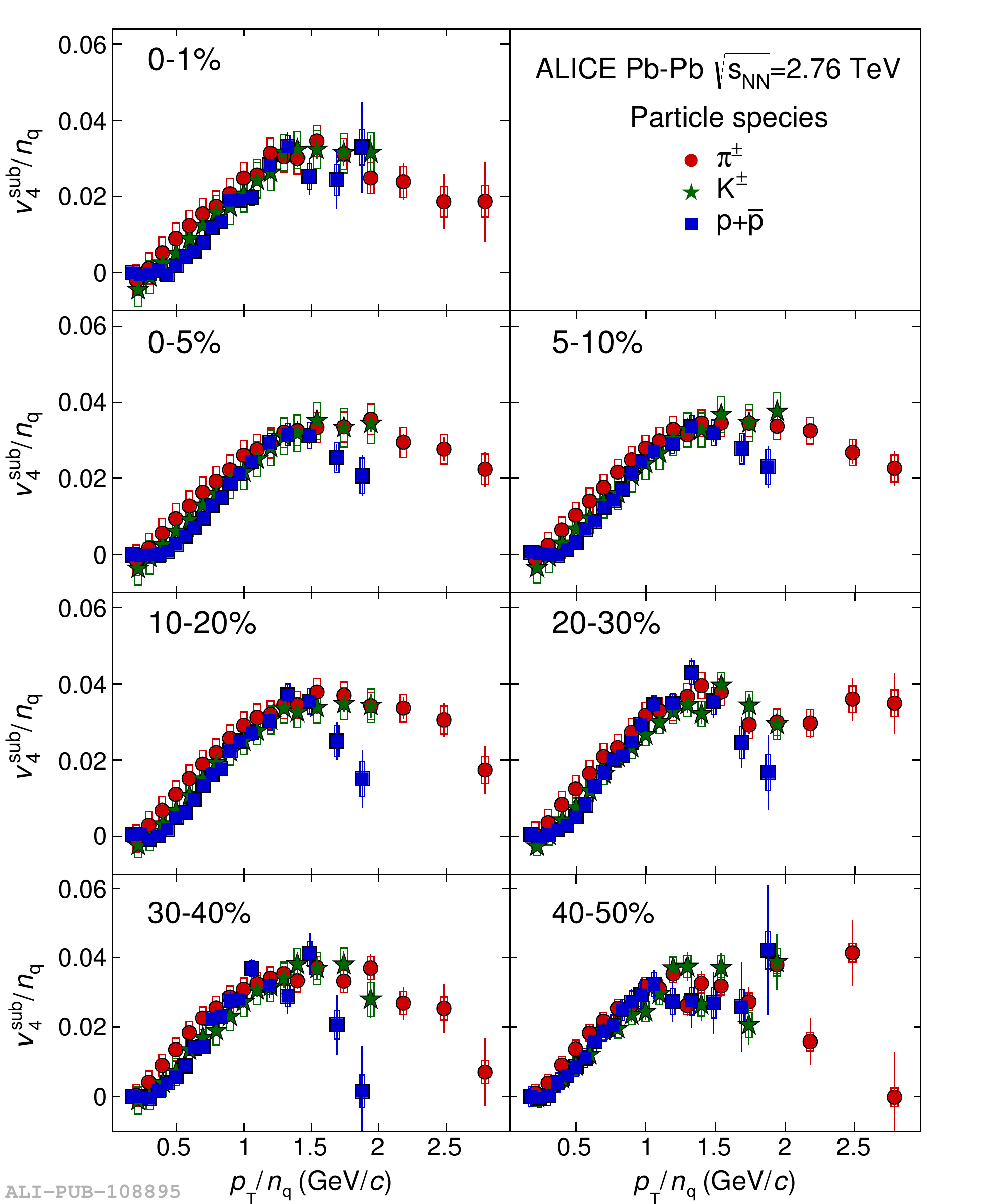}
\includegraphics[width=6.3cm,clip]{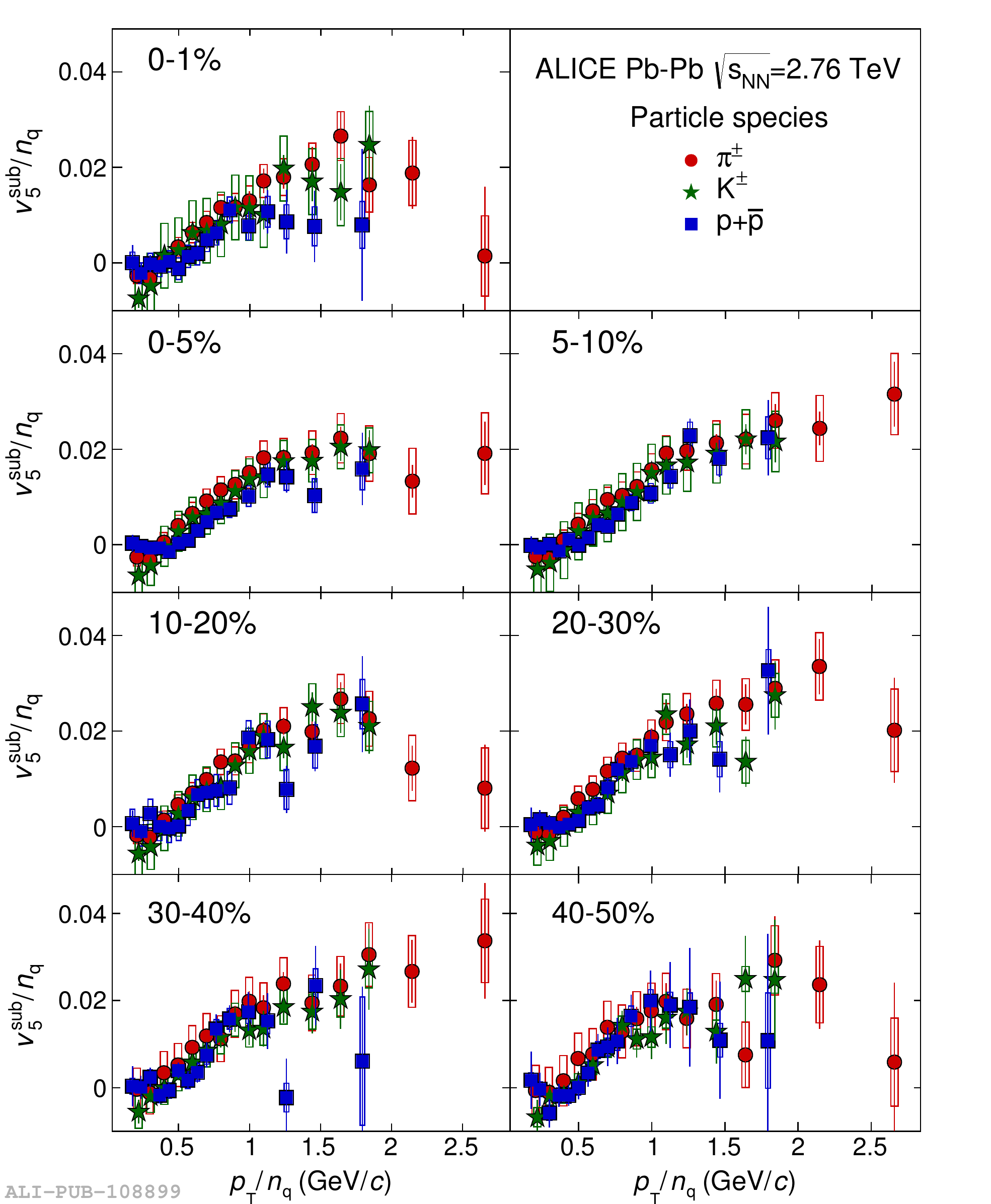}
\caption{$v_{2}^{\rm sub}/n_{q}$ (top left), $v_{3}^{\rm sub}/n_{q}$ (top right), $v_{4}^{\rm sub}/n_{q}$ (bottom left) and $v_{5}^{\rm sub}/n_{q}$ (bottom right) as a function of $p_{\rm T}/n_{q}$ for identified particles in various centrality classes in Pb--Pb collisions at $\sqrt{s_{\rm NN}} =$ 2.76 TeV. Figure is taken from~\cite{Adam:2016nfo}.}
\label{fig-22}       
\end{figure}

To better extract information on dynamic of system evolution, especially the information of freeze-out conditions, the anisotropic flow of identified particles are compared to hydrodynamic calculations based on iEBE-VISHNU~
\cite{Xu:2016hmp}. This is an event-by-event based hybrid model, which uses AMPT as initial conditions and couples 2$+$1 dimension viscous hydrodynamic calculations to a hadronic cascade model UrQMD. Figure~\ref{fig-21} shows that iEBE-VISHNU model could successfully produce the mass ordering feature of anisotropic flow for identified particles. More specifically, the model over-predicts $v_{2}$ in average by a 10 \% for pion, 10-15 \% for kaons and compatible with protons in the level of approximation of 10 \%. The agreements with data seems become better in more peripheral collisions. For higher harmonic flow $v_{3}$ and $v_{4}$, the model describe the data with a reasonable accuracy, they are consistent within 5\% for $\pt <$ 2 GeV/$c$ in presented centrality classes.

It was observed at RHIC that the $v_2$ of baryons is larger than that of mesons at the intermediate $\pt$ region (roughly $2<\pt<4$ GeV/$c$)~\cite{Adams:2003am, Abelev:2007qg, Adler:2003kt}. It was suggested that if both $v_2$ and $\pt$ are scaled by the number of constituent quarks ($n_q$), an universe scaling was expected for the $v_{2}/n_{q}$ as a function of $\pt/n_{q}$ for different particle species. This so-called number of constituent quark scaling (NCQ) was observed at RHIC with exceptional of pion $v_2$~\cite{Adare:2012vq}, but holds at an approximate level of $\pm 20$\% for $v_2$ at the LHC~\cite{Abelev:2014pua}. Such scaling is explained by the fact that the quark coalescence might be the dominant particle production mechanism in the intermediate $\pt$ region.
Figure~\ref{fig-22} presents the NCQ scaling for $v_{2}$, $v_{3}$, $v_{4}$ and $v_{5}$ for charged pions, kaons and protons in various centrality classes. It was observed that the NCQ scaling holds at the level of $\pm 20$ \% within the current statistical and systematic uncertainties, not only for elliptic flow which reported before in~\cite{Abelev:2014pua}, but also for higher harmonic flow.

\section{Correlations of anisotropic flow harmonics}
\label{sec-3}

Recently, it is proposed that the correlations between different order flow harmonics have unique sensitivities to initial conditions and $\eta/s$ of the produced matter in hydrodynamic calculations. This was firstly investigated by ATLAS Collaboration using so-called Event Shape Engineering (ESE)~\cite{Aad:2015lwa}. Later on, a new observable named Symmetric 2-harmonic 4-particle Cumulants, $SC(m,n)$, was proposed to quantitatively study the relationship between different orders of anisotropic flow harmonics $v_m$ and $v_n$~\cite{Bilandzic:2013kga}. It is defined as:
\begin{equation}
\begin{aligned}
SC(m,n) &~ = \left< \left< \cos (m\phi_{1} + n\phi_2 - m\phi_3 - n\phi_4) \right> \right>  -  \left< \left< \cos (m\phi_{1} - m\phi_2) \right> \right> \,\left< \left< \cos (n\phi_{1} - n\phi_2) \right> \right> \\
&~ = \langle v_m^2  \,  v_n^2 \rangle -  \langle v_m^2  \rangle \, \langle v_n^2 \rangle,
\end{aligned}
\end{equation}
with $n \neq m$. 
By construction, this observable should be insensitive to non-flow effects due to usage of 4-particle cumulant and it is independent on the correlations between various symmetry planes. Thus, it is non-zero only if there is correlation (or anti-correlation) between $v_m$ and $v_n$. Previous AMPT model calculations showed that $SC(3,2)$ is negative (anti-correlations between $v_{2}$ and $v_{3}$) while $SC(4,2)$ is positive (correlations between $v_{2}$ and $v_{4}$). The observed (anti-)correlations are also sensitive to the transport properties of the created system, e.g. the partonic and hadronic interactions~\cite{Zhou:2015eya, Bilandzic:2013kga}.

\begin{figure}[h]
\centering
\includegraphics[width=14cm,clip]{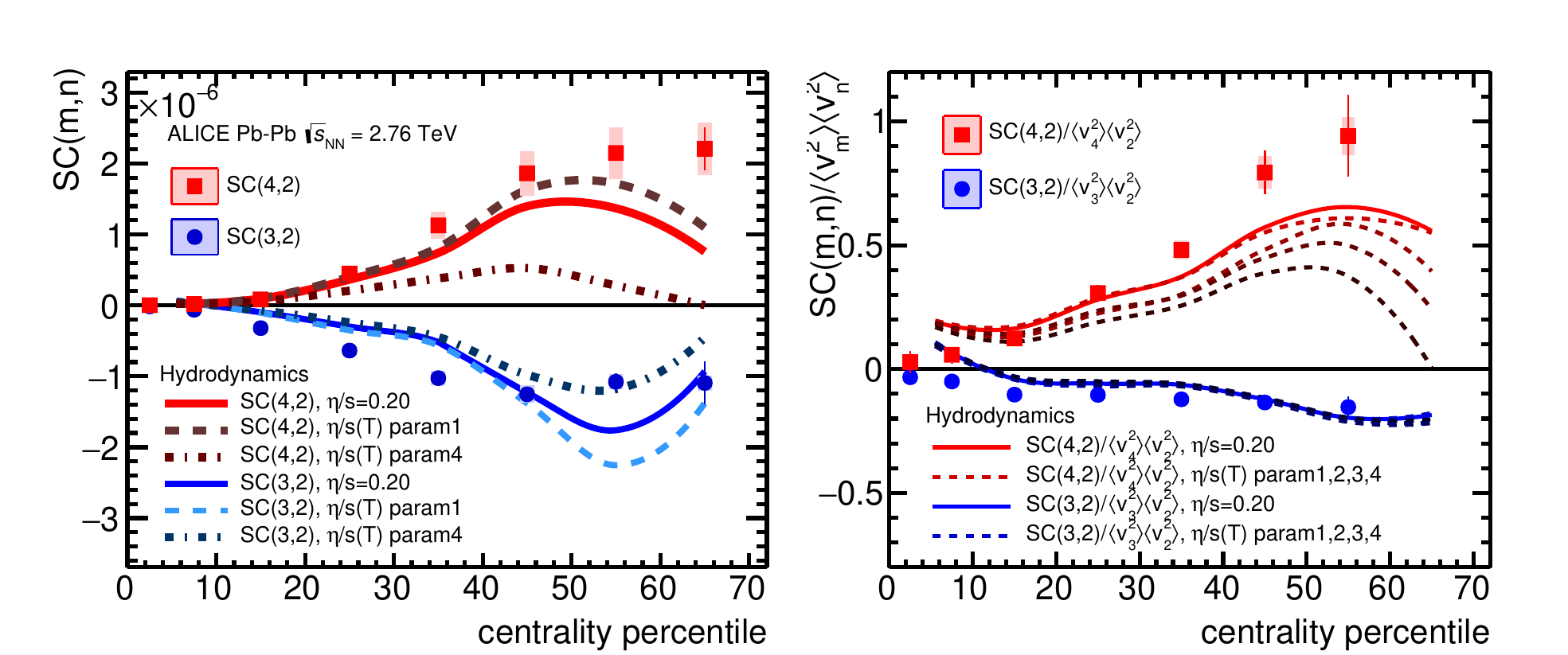}
\caption{The centrality dependence of $SC(m,n)$ (left) and $NSC(m,n)$ (right) at $\sqrt{s_{_{\rm NN}}}$ = 2.76 TeV Pb--Pb collisions, the hydrodynamic calculations from~\cite{Niemi:2015qia} are presented for comparisons.}
\label{fig:sc_ALICE}       
\end{figure}

The first ALICE measurements of centrality dependence of $SC(4,2)$ (red squares) and $SC(3,2)$ (blue circles) are presented in Fig.~\ref{fig:sc_ALICE} (left). Positive values of $SC(4,2)$ and negative results of $SC(3,2)$ are observed for all centralities, as correctly predicted by AMPT~\cite{Bilandzic:2013kga}. This confirms a correlation between the event-by-event fluctuations of $v_{2}$ and $v_{4}$, and an anti-correlation between $v_{2}$ and $v_{3}$.
The like-sign technique, which is a powerful approach to suppress non-flow effects~\cite{Aamodt:2010pa}, is applied. It is found that the difference between correlations for like-sign and all charged combinations are much smaller compared to the magnitudes of $SC(m,n)$ itself. Therefore, the non-zero values of $SC(m,n)$ presented here cannot be explained by non-flow effects solely.
To further study the properties of QGP, we show in Fig.~\ref{fig:sc_ALICE} (right) the comparison of $SC(m,n)$ measurements and hydrodynamic model calculations. This calculation is from the event-by-event perturbative-QCD$+$saturation$+$hydrodynamic (``EKRT") framework~\cite{Niemi:2015qia}. It incorporates both initial conditions and hydrodynamic evolution. It is observed that $SC(m,n)$ is very sensitive to the values of $\eta/s$ in this model. Thus, the investigation of $SC(m,n)$ provides a new approach to constrain the input of $\eta/s$ in hydrodynamic calculations, in addition to standard anisotropic flow studies.
We also notice that although this hydrodynamic model reproduced the centrality dependence of $v_n$ for $n \leq 2$ fairly well using various $\eta/s$(T) parametrization~\cite{Niemi:2015qia}, its predictions of $SC(m,n)$ cannot quantitatively describe the data. There is no single centrality for which a given $\eta/s$(T) parameterization describes simultaneously $SC(4,2)$ and $SC(3,2)$. Therefore, it is believed that the new observable $SC(m,n)$ provides stronger constrains on the $\eta/s$ and initial conditions than individual $v_n$ measurement alone.

In order to further investigate the correlations between $v_{m}$ and $v_{n}$ without contributions from flow harmonics, it is proposed to divide $SC(m,n)$ by the products $\langle v_m^2\rangle \langle v_n^2\rangle$. This new observable is named $NSC(m,n)$. The measurements from ALICE are presented in Fig.~\ref{fig:sc_ALICE}~(right). 
Here 2-particle correlations $\langle v_m^2 \rangle$ and $\langle v_n^2\rangle$ are obtained with a pseudorapidity gap of $|\Delta\eta| > 1.0$ to suppress short range non-flow effects. 
Similar to $SC(4,2)$, the normalized $SC(4,2)$ observable still exhibits a significant sensitivity to different $\eta/s$ parameterizations and the initial conditions, which again provides a good opportunity to discriminate between various possibilities of the detailed setting of $\eta/s({\rm T})$ of the produced QGP and the initial conditions used in hydrodynamic calculations. However, $NSC(3,2)$ is independent of the setting of $\eta/s({\rm T})$. In addition, it was demonstrated in~\cite{Zhu:2016puf} that $NSC(3,2)$ is compatible with its corresponding observable $NSC^{\varepsilon}(3,2)$ in the initial state. Thus, the $NSC^{v}(3,2)$ measurements could be taken as a crucial observable to constrain initial conditions without concern of transport properties of the system. 
Furthermore, none of existing theoretical calculations can reproduce the data, future developments of $SC(m,n)$ and $NSC(m,n)$ involved higher harmonics and more particle correlations~\cite{Zhou:2016eiz}, together with comparisons to further tuned hydrodynamic calculations, are crucial for deep understanding of correlations between different oder flow harmonics.

\section{Anisotropic flow at $\sqrt{s_{\rm NN}} = 5.02$ TeV}
\label{sec-3}

\begin{figure}[h]
\centering
\includegraphics[width=7cm,clip]{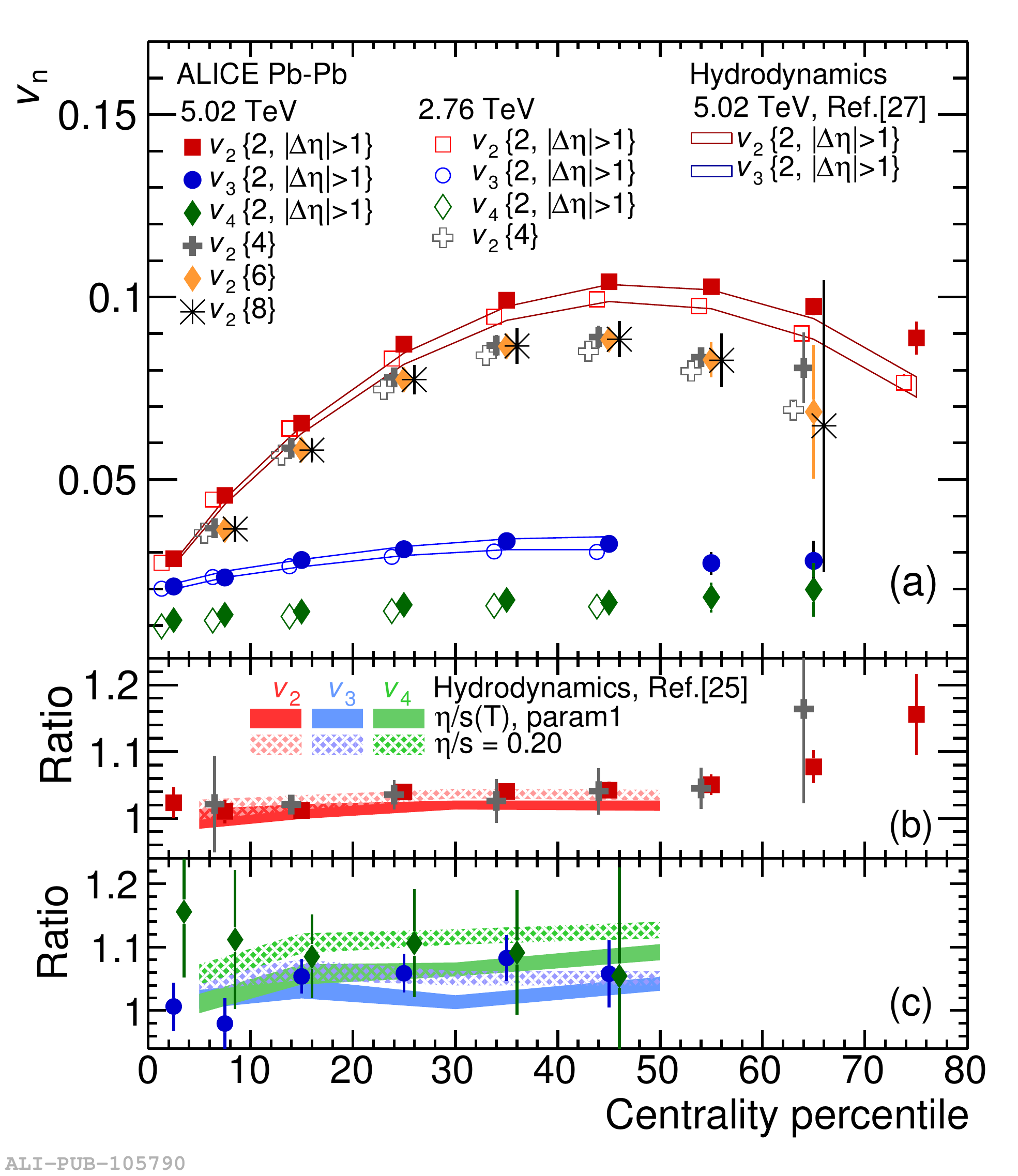}
\includegraphics[width=7cm,clip]{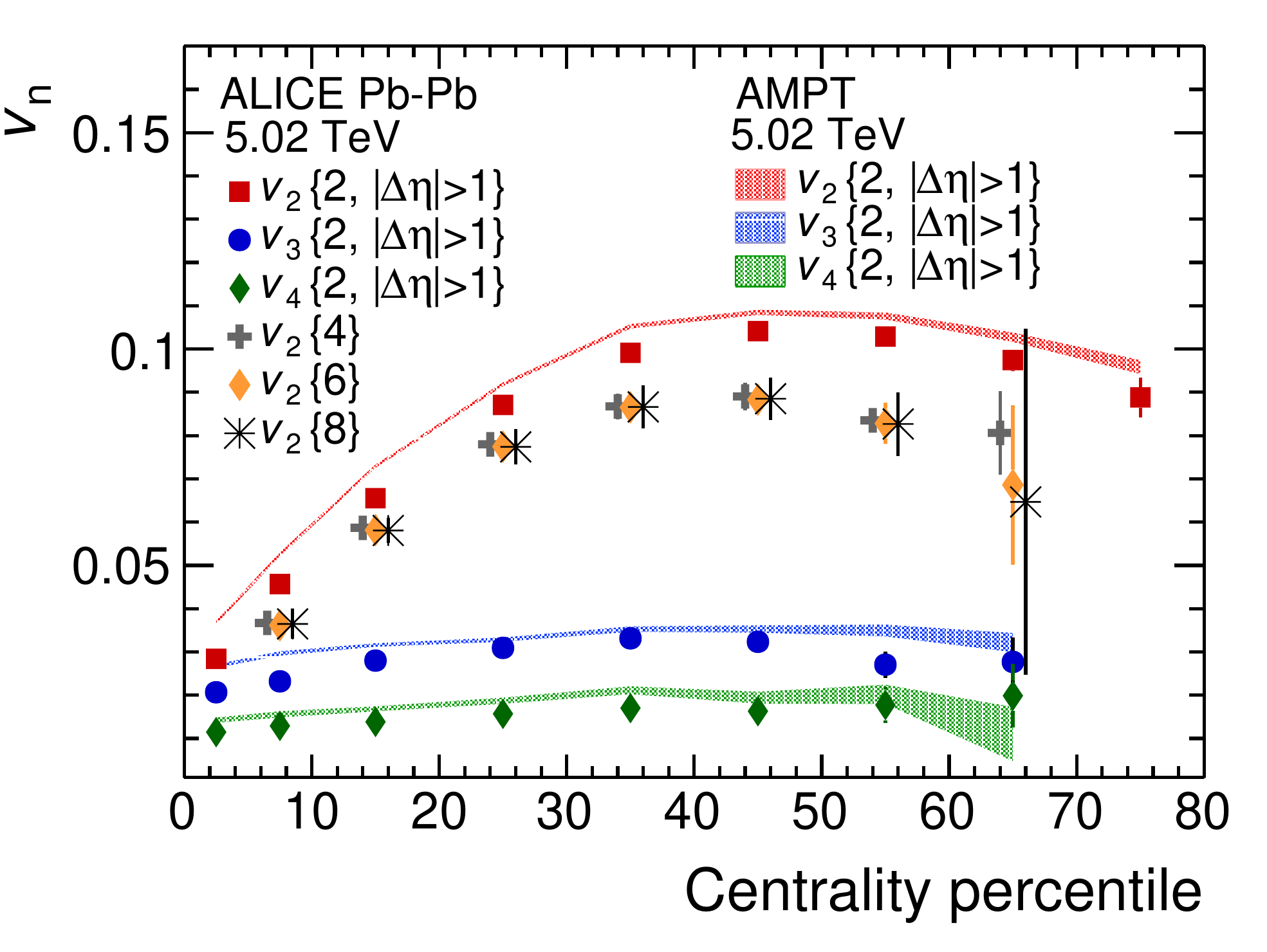}
\caption{Anisotropic flow as a function of centrality. Hydrodynamic~\cite{Noronha-Hostler:2015uye, Niemi:2015voa} and AMPT~\cite{Feng:2016emh} calculations are presented for comparisons.}
\label{fig-run2flow}      
\end{figure}

Exploiting the data collected during November 2015 with Pb--Pb collisions at record breaking energies of $\sqrt{s_{_{\rm NN}}}=5.02$~TeV, ALICE measured for the first time $v_{2}$, $v_{3}$ and $v_{4}$ of charged particles at this energy~\cite{Adam:2016izf}.
Centrality dependence of anisotropic flow using two-particle correlations with pseudorapidity and multi-particle cumulants methods are shown in Fig.~\ref{fig-run2flow}. It is observed that $v_{2}$ increases from central to peripheral collisions, saturates in the 40--50\% centrality class with a maximum value about 0.10 for two-particle correlations and is around 0.09 for multi-particle cumulants. The multi-particle cumulants show similar centrality dependence and the results from four-, six- and eight-particle cumulants are consistent with each other.
On the other hand, the significant differences between two- and multi-particle cumulants could be attribute to different sensitivities to elliptic flow fluctuations. The higher harmonic flow $v_3$ and $v_4$ are smaller and their centrality dependence is much weaker compared to $v_{2}$.
In addition, anisotropic flow measurements are compared to hydrodynamic calculations which made a robust prediction without relying on a particular model for initial conditions and without precise knowledge of medium properties such as viscosity~\cite{Noronha-Hostler:2015uye}. As shown in Fig.~\ref{fig-run2flow} left (a), the hydrodynamic calculation are compatible with the measured anisotropic flow $v_n$. Furthermore, the AMPT model based on a new set of input parameters only slightly overestimates the anisotropic flow measurements~\cite{Feng:2016emh}, as shown in Fig.~\ref{fig-run2flow} (right).

Figure~\ref{fig-run2flow} left (b) and (c) show the ratios of $v_n$ measured at 5.02 TeV to 2.76 TeV.
The increase of $v_{2}$ and $v_{3}$ is rather moderate from the two energies, with an increase of (3.0$\pm$0.6)\%, (4.3$\pm$1.4)\%, respectively. The increase is more significant for $v_{4}$, which is (10.2$\pm$3.8)\%, over the centrality range 0--50\% in Pb--Pb collisions.
This increase of anisotropic flow is compatible with hydrodynamic predictions~\cite{Niemi:2015voa}. The data seems support a low value of $\eta/s$ for the created matter in Pb--Pb collisions at $\sqrt{s_{_{\rm NN}}} = $ 5.02~TeV, suggests that the $\eta/s$ does not change significantly with respect to the one at lower energy.

\begin{figure}[h]
\centering
\includegraphics[width=7cm,clip]{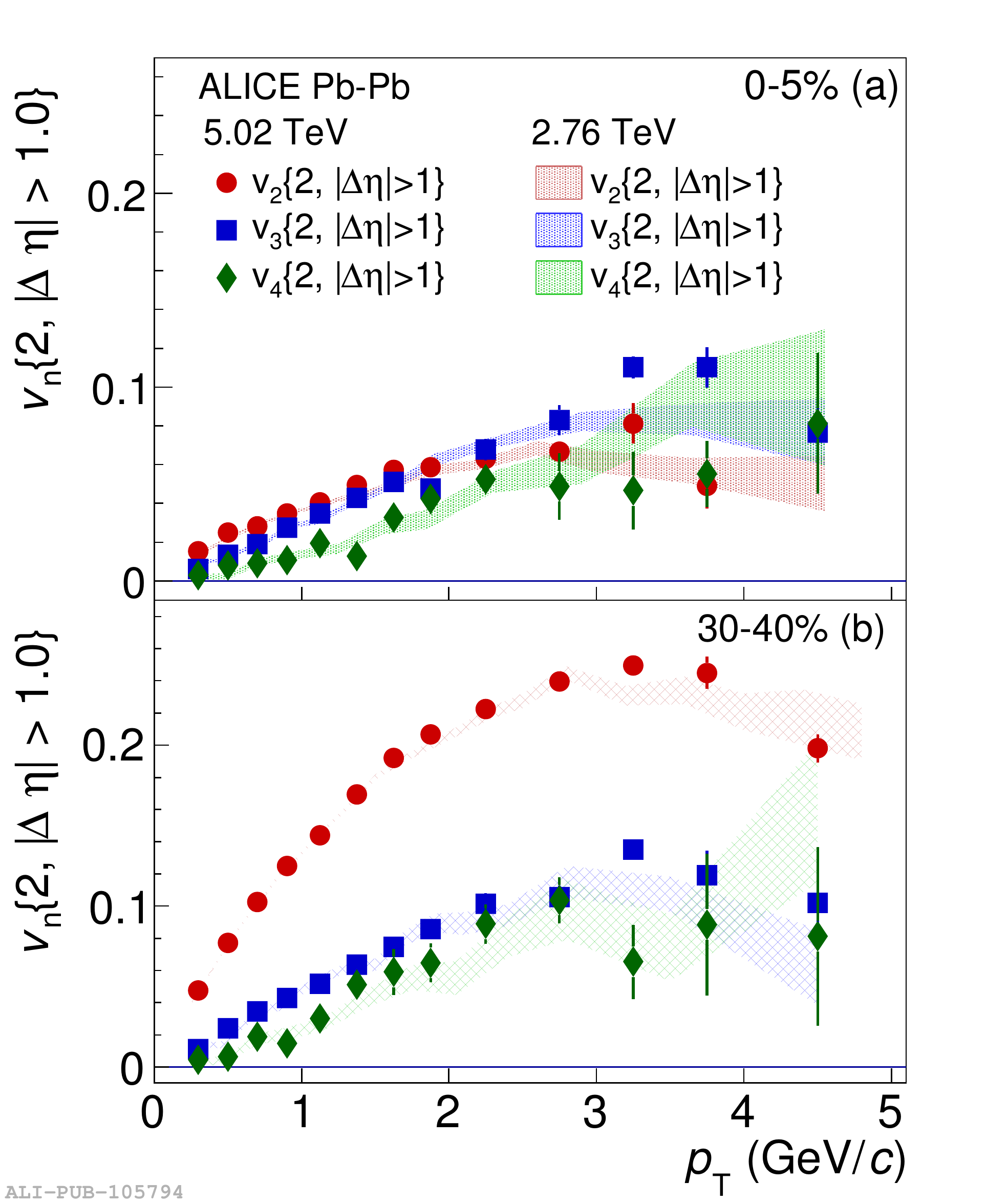}
\includegraphics[width=7cm,clip]{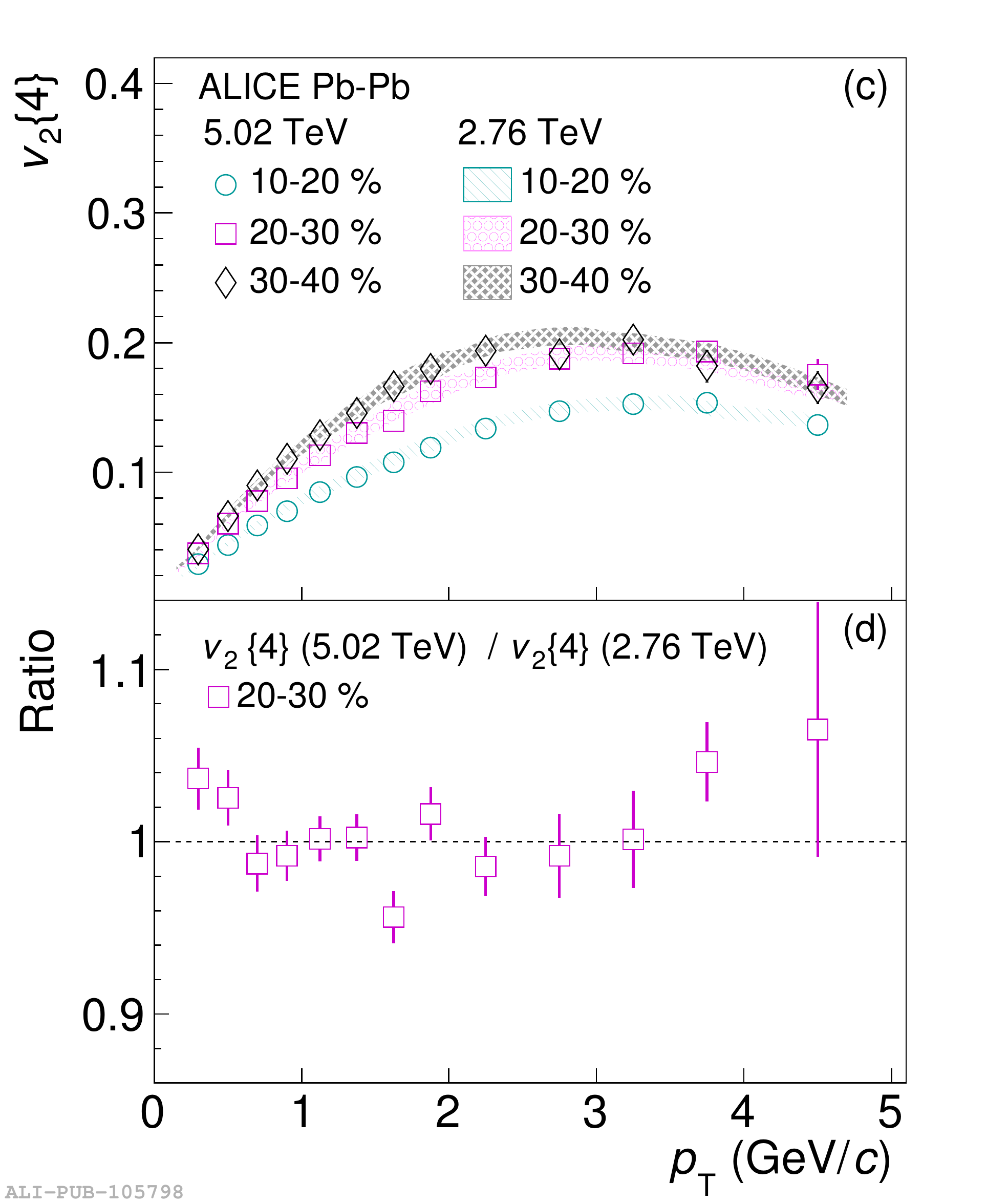}
\caption{The $p_{\rm T}$-differential flow measured in selected centrality classes. The same measurements from 2.76 TeV are presented for comparisons.}
\label{fig-run2ptflow}       
\end{figure}

The transverse momentum ($p_{\rm T}$) dependent anisotropic flow $v_{n}$ ($n=2,3,4$) are presented for selected centrality classes. The results are consistent with the previous $v_{n}(p_{\rm T})$ measurements from 2.76 TeV, as shown in Fig.~\ref{fig-run2ptflow}. It suggests that the observed increase of $p_{\rm T}$ integrated flow results seen in Fig.~\ref{fig-run2flow} could be mainly attributed to stronger radial flow produced at 5.02 TeV. This stronger radial flow results in an increase of mean transverse momentum $\langle p_{\rm T} \rangle$ from 2.76 to 5.02 TeV. Future comparisons of $p_{\rm T}$-differential flow between experimental measurements and hydrodynamic calculations will provide important information to constrain further details of the theoretical calculations, e.g. determination of radial flow and freeze-out conditions.

\section*{Acknowledgments}

This work is supported by the Danish Council for Independent Research, Natural Sciences, and the Danish National Research Foundation (Danmarks Grundforskningsfond).



\begin{thebibliography}{50}

\bibitem{Lee:1978mf}
T.D. Lee, Phys. Rev. \textbf{D19}, 1802 (1979), [,174(1978)]

\bibitem{Shuryak:1980tp}
E.V. Shuryak, Phys. Rept. \textbf{61}, 71 (1980)

\bibitem{Ollitrault:1992bk}
J.Y. Ollitrault, Phys. Rev. \textbf{D46}, 229 (1992)

\bibitem{Voloshin:1994mz}
S.~Voloshin, Y.~Zhang, Z. Phys. \textbf{C70}, 665 (1996),
  \texttt{hep-ph/9407282}

\bibitem{Ackermann:2000tr}
K.H. Ackermann et~al. (STAR), Phys. Rev. Lett. \textbf{86}, 402 (2001),
  \texttt{nucl-ex/0009011}

\bibitem{Adcox:2002ms}
K.~Adcox et~al. (PHENIX), Phys. Rev. Lett. \textbf{89}, 212301 (2002),
  \texttt{nucl-ex/0204005}

\bibitem{Back:2002gz}
B.B. Back et~al. (PHOBOS), Phys. Rev. Lett. \textbf{89}, 222301 (2002),
  \texttt{nucl-ex/0205021}

\bibitem{Huovinen:2001cy}
P.~Huovinen, P.F. Kolb, U.W. Heinz, P.V. Ruuskanen, S.A. Voloshin, Phys. Lett.
  \textbf{B503}, 58 (2001), \texttt{hep-ph/0101136}

\bibitem{Kolb:2000fha}
P.F. Kolb, P.~Huovinen, U.W. Heinz, H.~Heiselberg, Phys. Lett. \textbf{B500},
  232 (2001), \texttt{hep-ph/0012137}

\bibitem{Luzum:2008cw}
M.~Luzum, P.~Romatschke, Phys. Rev. \textbf{C78}, 034915 (2008), [Erratum:
  Phys. Rev.C79,039903(2009)], \texttt{0804.4015}

\bibitem{Song:2007ux}
H.~Song, U.W. Heinz, Phys. Rev. \textbf{C77}, 064901 (2008), \texttt{0712.3715}

\bibitem{Song:2010mg}
H.~Song, S.A. Bass, U.~Heinz, T.~Hirano, C.~Shen, Phys. Rev. Lett.
  \textbf{106}, 192301 (2011), [Erratum: Phys. Rev. Lett.109,139904(2012)],
  \texttt{1011.2783}

\bibitem{Aamodt:2010pa}
K.~Aamodt et~al. (ALICE), Phys. Rev. Lett. \textbf{105}, 252302 (2010),
  \texttt{1011.3914}

\bibitem{ALICE:2011ab}
K.~Aamodt et~al. (ALICE), Phys. Rev. Lett. \textbf{107}, 032301 (2011),
  \texttt{1105.3865}

\bibitem{ATLAS:2011ah}
G.~Aad et~al. (ATLAS), Phys. Lett. \textbf{B707}, 330 (2012),
  \texttt{1108.6018}

\bibitem{Chatrchyan:2012wg}
S.~Chatrchyan et~al. (CMS), Eur. Phys. J. \textbf{C72}, 2012 (2012),
  \texttt{1201.3158}

\bibitem{Heinz:2013th}
U.~Heinz, R.~Snellings, Ann. Rev. Nucl. Part. Sci. \textbf{63}, 123 (2013),
  \texttt{1301.2826}

\bibitem{Luzum:2013yya}
M.~Luzum, H.~Petersen, J. Phys. \textbf{G41}, 063102 (2014), \texttt{1312.5503}

\bibitem{Huovinen:2013wma}
P.~Huovinen, Int. J. Mod. Phys. \textbf{E22}, 1330029 (2013),
  \texttt{1311.1849}

\bibitem{Shuryak:2014zxa}
E.~Shuryak (2014), \texttt{1412.8393}

\bibitem{Song:2013gia}
H.~Song, Pramana \textbf{84}, 703 (2015), \texttt{1401.0079}

\bibitem{Dusling:2015gta}
K.~Dusling, W.~Li, B.~Schenke, Int. J. Mod. Phys. \textbf{E25}, 1630002 (2016),
  \texttt{1509.07939}

\bibitem{Kovtun:2004de}
P.~Kovtun, D.T. Son, A.O. Starinets, Phys. Rev. Lett. \textbf{94}, 111601
  (2005), \texttt{hep-th/0405231}

\bibitem{Adam:2016ows}
J.~Adam et~al. (ALICE) (2016), \texttt{1605.02035}

\bibitem{Adam:2016nfo}
J.~Adam et~al. (ALICE), JHEP \textbf{09}, 164 (2016), \texttt{1606.06057}

\bibitem{ALICE:2016kpq}
J.~Adam et~al. (ALICE) (2016), \texttt{1604.07663}

\bibitem{Adam:2016izf}
J.~Adam et~al. (ALICE), Phys. Rev. Lett. \textbf{116}, 132302 (2016),
  \texttt{1602.01119}

\bibitem{Alver:2010gr}
B.~Alver, G.~Roland, Phys. Rev. \textbf{C81}, 054905 (2010), [Erratum: Phys.
  Rev.C82,039903(2010)], \texttt{1003.0194}

\bibitem{Back:2004zg}
B.B. Back et~al. (PHOBOS), Phys. Rev. Lett. \textbf{94}, 122303 (2005),
  \texttt{nucl-ex/0406021}

\bibitem{Alver:2006wh}
B.~Alver et~al. (PHOBOS), Phys. Rev. Lett. \textbf{98}, 242302 (2007),
  \texttt{nucl-ex/0610037}

\bibitem{Bearden:2001qq}
I.G. Bearden et~al. (BRAHMS), Phys. Rev. Lett. \textbf{88}, 202301 (2002),
  \texttt{nucl-ex/0112001}

\bibitem{Denicol:2015nhu}
G.~Denicol, A.~Monnai, B.~Schenke, Phys. Rev. Lett. \textbf{116}, 212301
  (2016), \texttt{1512.01538}

\bibitem{Xu:2011fi}
J.~Xu, C.M. Ko, Phys. Rev. \textbf{C83}, 034904 (2011), \texttt{1101.2231}

\bibitem{Aad:2014eoa}
G.~Aad et~al. (ATLAS), Eur. Phys. J. \textbf{C74}, 2982 (2014),
  \texttt{1405.3936}

\bibitem{Chatrchyan:2012ta}
S.~Chatrchyan et~al. (CMS), Phys. Rev. \textbf{C87}, 014902 (2013),
  \texttt{1204.1409}

\bibitem{Xu:2016hmp}
H.j. Xu, Z.~Li, H.~Song, Phys. Rev. \textbf{C93}, 064905 (2016),
  \texttt{1602.02029}

\bibitem{Abelev:2014pua}
B.B. Abelev et~al. (ALICE), JHEP \textbf{06}, 190 (2015), \texttt{1405.4632}

\bibitem{Adams:2003am}
J.~Adams et~al. (STAR), Phys. Rev. Lett. \textbf{92}, 052302 (2004),
  \texttt{nucl-ex/0306007}

\bibitem{Abelev:2007qg}
B.I. Abelev et~al. (STAR), Phys. Rev. \textbf{C75}, 054906 (2007),
  \texttt{nucl-ex/0701010}

\bibitem{Adler:2003kt}
S.S. Adler et~al. (PHENIX), Phys. Rev. Lett. \textbf{91}, 182301 (2003),
  \texttt{nucl-ex/0305013}

\bibitem{Adare:2012vq}
A.~Adare et~al. (PHENIX), Phys. Rev. \textbf{C85}, 064914 (2012),
  \texttt{1203.2644}

\bibitem{Aad:2015lwa}
G.~Aad et~al. (ATLAS), Phys. Rev. \textbf{C92}, 034903 (2015),
  \texttt{1504.01289}

\bibitem{Bilandzic:2013kga}
A.~Bilandzic, C.H. Christensen, K.~Gulbrandsen, A.~Hansen, Y.~Zhou, Phys. Rev.
  \textbf{C89}, 064904 (2014), \texttt{1312.3572}

\bibitem{Zhou:2015eya}
Y.~Zhou, K.~Xiao, Z.~Feng, F.~Liu, R.~Snellings (2015), \texttt{1508.03306}

\bibitem{Niemi:2015qia}
H.~Niemi, K.J. Eskola, R.~Paatelainen (2015), \texttt{1505.02677}

\bibitem{Zhu:2016puf}
X.~Zhu, Y.~Zhou, H.~Xu, H.~Song (2016), \texttt{1608.05305}

\bibitem{Zhou:2016eiz}
Y.~Zhou, Adv. High Energy Phys. \textbf{2016}, 9365637 (2016),
  \texttt{1607.05613}

\bibitem{Noronha-Hostler:2015uye}
J.~Noronha-Hostler, M.~Luzum, J.Y. Ollitrault, Phys. Rev. \textbf{C93}, 034912
  (2016), \texttt{1511.06289}

\bibitem{Niemi:2015voa}
H.~Niemi, K.J. Eskola, R.~Paatelainen, K.~Tuominen, Phys. Rev. \textbf{C93},
  014912 (2016), \texttt{1511.04296}

\bibitem{Feng:2016emh}
Z.~Feng, G.M. Huang, F.~Liu (2016), \texttt{1606.02416}

\end{thebibliography}
\end{document}